\documentstyle[12pt]{article}
\font\sf=cmss10                    %San-Serif 10
\makeatletter
\@addtoreset{equation}{section}

\makeatother
\input BoxedEPS.tex
\SetRokickiEPSFSpecial
\HideDisplacementBoxes

\def\del{\partial}
\def\Dslash{\not{\hbox{\kern-4pt $D$}}}
\def\dslash{\not{\hbox{\kern-2pt $\del$}}}
\def\gslash{\not{\hbox{\kern-2pt $\gamma$}}}
\def\g5{\gamma_5}

\def\half{\frac{1}{2}}

\def\ra{\rightarrow}

\def\ch#1#2{({#1 \atop #2 })}
\newcommand{\mathbold}[1]{\mbox{\boldmath $\bf#1$}}
\def\mJ{\mathbold{J}}
\def\malpha{\mathbold{\alpha}}
\def\ms{\mathbold{s}}
\def\ma{\mathbold{a}}
\def\mb{\mathbold{b}}
\def\mc{\mathbold{c}}
\def\momega{\mathbold{\omega}}
\def\meta{\mathbold{\eta}}
\def\bbbz {{\sf Z\!\!\!Z}}
\def\sl2z{SL(2,\bbbz)}
\def\fracs#1#2{\textstyle\frac #1#2}

\begin{document}

{\flushright{\small MIT-CTP-2730\\
hep-th/9804210\\
%\today \\
}}

\vspace{1.0in}
\begin{center}\LARGE
{\bf String Junctions for Arbitrary Lie Algebra Representations  \\}
\end{center} \vskip 0.8cm
\begin{center}
{\large Oliver DeWolfe\footnote{\noindent E-mail: 
odewolfe@ctp.mit.edu} and Barton Zwiebach\footnote{\noindent 
E-mail: 
zwiebach@irene.mit.edu \\
\hspace*{.15in} Work supported by the U.S.\ Department of Energy 
under contract \#DE-FC02-94ER40818.}}
\vskip 0.2cm
{{\it Center for Theoretical Physics,\\
Laboratory for Nuclear Science,\\
Department of Physics\\
Massachusetts Institute of Technology\\
Cambridge, Massachusetts 02139, U.S.A.}}

\end{center}

\vspace{0.5in}
\begin{abstract} 
We consider string junctions with endpoints on a set of
branes of IIB string theory defining
an $ADE$-type gauge Lie algebra.  
We show how to characterize uniquely equivalence classes
of junctions related by string/brane crossing through invariant charges that 
count the effective number of prongs ending on each brane.
Each equivalence class defines a point on a lattice of junctions.
We define a metric on this lattice arising from the intersection 
pairing of junctions, and use self-intersection to identify 
junctions in the adjoint and fundamental representations of all $ADE$ algebras.
This information suffices to determine the relation between 
junction lattices and the Lie-algebra weight lattices. 
Arbitrary representations are built by allowing junctions with asymptotic
$(p,q)$ charges, on which the group of conjugacy classes of representations
is represented additively. One can
view the $(p,q)$ asymptotic charges as Dynkin labels associated to two
new fundamental weight vectors.  
\end{abstract}

\newpage

\tableofcontents

\listoffigures

\section{Introduction and summary}\label{s:1}

In the modern viewpoint on gauge symmetry enhancement
in open string theory, the gauge vectors associated
to a $u(n)$ gauge algebra are seen to arise from 
open strings beginning and ending on 
a collection of $n$ Dirichlet branes (see \cite{polchinski}).
Given the utility and simplicity of brane pictures
for gauge theory it is of interest to understand the
brane picture of other gauge algebras. A natural laboratory
for this problem is IIB superstrings compactified on
a two-sphere in the presence of several 7-branes appearing
as points on this sphere and extending along the other spatial
dimensions.  In this background 
 both strings and seven branes carry $(p,q)$ labels.
Here $p$ and $q$ are relatively prime integers denoting
for the case of a string the charges under the NS-NS and
RR antisymmetric tensors respectively. A $[p,q]$ seven-brane  
is a seven-brane where a $({p\atop q})$ string can end.  
The F-theory picture \cite{vafa} of this superstring background
was used to obtain the specific 7-brane content that would
lead to the $D_n$ ($so(2n)$) and $E_n$ ($n=6,7,8$) gauge algebras
\cite{sen, dasgupta, johansen}. Such gauge
algebras require branes of various $[p,q]$ types. 

It was naturally assumed that open strings, represented by
geodesic lines on the two sphere joining suitable branes, 
would give rise to the gauge vectors of the corresponding 
algebra. This is not always 
the case; depending on the positions
of the branes, a given gauge vector is realized sometimes
as a geodesic string, but more generally it is realized as a
geodesic string junction \cite{GZ}. Indeed, such transitions
were examined explicitly in \cite{mgthbz,hauer} where strong evidence
was given to the effect that given a specific configuration
of branes either a unique open string or a unique string junction is
geodesic. 
String junctions, originally introduced in \cite{aharony, schwarzreview},
have been the subject of much recent study with regards to their
BPS properties \cite{dasguptamukhi, sennetwork, krogh, matsuo, kishimoto},
their dynamics \cite{rey, callan}, 
their role in ${\cal N}=4$ super-Yang-Mills \cite{OB,hata}, as well
as for their relevance to dualities \cite{imamura} and to the analysis of
BPS states of ${\cal N}=2$ SYM theory  \cite{bf,nekrasov}. 
There is also a parallel line of development dealing with $[p,q]$ five-branes 
and five dimensional gauge theory \cite{AHK,leungvafa}.
   
The main motivation for considering junctions in \cite{GZ} was
that such string states can carry manifestly the 
gauge charges under the obvious subalgebras and this enables one 
to reproduce the  structure of the exceptional algebras by combining junctions the
same way one combines open strings. The purpose of the present paper is to
give a firm Lie-algebraic basis to junctions, explaining the reason
for much of the coincidences noted in \cite{GZ}. While that work  dealt with
adjoint representations only, we will be able to discuss here arbitrary 
representations of the {\it ADE} series of algebras. Since junctions represent states
with gauge charges, they are naturally associated to weight vectors belonging 
to some representation of the relevant algebra. The main goal of the present
paper is to show how to calculate the weight vector associated with
an arbitrary junction, and conversely, given an arbitrary representation 
of an {\it ADE} algebra, to show how to construct the junctions representing
the corresponding string states.   
It is clear that representations other than adjoints are present
in applications, typically as massive BPS states \cite{iqbal}. Some of our
results related to $so(8)$ symmetry were anticipated
in an interesting paper by Imamura \cite{imamura}, who among other 
things showed  how to construct the ${\bf 8_v,8_s}$ and ${\bf 8_c}$
representations. Our treatment will focus only on the Lie-algebraic issues, and we
will not discuss whether the general junctions we consider have BPS
representatives.  This is an important issue and should be studied,
possibly along the lines of Mikhailov {\it et al.} \cite{nekrasov}.

Junctions fall into equivalence classes.  
Different members of
an equivalence class represent the same physical state in different regions
of the moduli space of backgrounds. The different members differ by
transformations that involve strings crossing branes, creating or destroying
prongs in this process \cite{GZ}. This process is U-dual to the original
version of the Hanany-Witten effect \cite{hanany,08}. In addition, junctions
that differ by smooth deformations 
are also to be considered equivalent, including the possibility of collapsing 
or resolving vertices. 
Since junctions with very different presentations can 
be equivalent, one must first learn how to tell when junctions are
equivalent.  We solve this problem by introducing 
a set of charges $Q^\mu$, one for   
each 7-brane of the background,  invariant under the operations that
preserve the equivalence class of a junction. These charges can be used to
construct a canonical presentation for any junction, making it
manifest that two junctions are equivalent {\it if and only if} their
invariant charges agree. For each brane, the invariant charge arises from two
sources, prongs ending on the brane and strings crossing the branch cut
emerging from the brane. A junction may be localized around the set of
seven-branes, or may have an asymptotic string (or strings) carrying $(p,q)$
charge away from the set of branes. 

\smallskip
The invariant charges allow us to think of junctions as linear combinations
with integer coefficients of a set of  basis strings $\ms_\mu$. 
For each brane, the associated basis string is a string departing the brane
and going off to infinity
without crossing any branch cut.  In this way the space of (equivalence classes
of) junctions naturally becomes a {\it lattice} over the set of basis strings.
This a very important concept since it immediately gives a complete
classification of all possible junctions, and helps make systematic 
any  search for special junctions. 
It is important
to emphasize that in each case the number of basis strings (or branes) exceeds
the rank of the Lie algebra to be generated by the background.  
This means that the basis of strings is not equivalent to a basis of junctions 
for the roots of the Lie algebra.
On the other hand, the sublattice of junctions with zero asymptotic charge is
in fact spanned by a number of junctions equal to the rank.

\smallskip
Motivated by the intersection invariant of 
homology two cycles in the F-theory (or M-theory) picture we introduce
a symmetric bilinear form  $(\cdot \: , \, \cdot)$ or metric on the junction
lattice. This is a slight modification of the standard concept, since our basis
strings are not homology cycles.  
With a little abuse of language, given a junction
$\mJ$ we call
$(\mJ , \mJ)$ the self-intersection number of the junction.\footnote{This
intersection bilinear coincides with the one used in the analysis of $su(2)$ 
super-Yang-Mills in Ref.~\cite{nekrasov}.}  
The metric is negative-definite for the $A_n$ series. It is degenerate for the
$D_n$ series; for the $D_4$ algebra there exist two null junctions (junctions
whose intersection with any junction vanish) while for $D_n$ algebras with
$n\geq 5$ there is one null junction. The metric is non-degenerate but of
indefinite signature for the
$E_n$ series; it has $n$ negative eigenvalues and two positive ones. 

\smallskip 
We  search the junction lattices for all inequivalent junctions  with zero
asymptotic charge and self-intersection
$(-2)$. For each case, we  find a set that is in correspondence with the 
set of roots of the corresponding Lie algebra. We suspect that this (simple!)
derivation of the Lie algebra from the brane configuration must
somehow parallel the work involved in understanding the 
detailed nature of the Kodaira singularities. 
All roots are presented as junctions, but, completing the work of
\cite{johansen,GZ} we show that they all have equivalent presentations as open 
strings without self-crossings (Jordan open strings).  Indeed any Jordan open
string with endpoints at two different branes has self-intersection $(-2)$,
but we are not aware of an argument that would imply that any junction of
self-intersection $(-2)$ must have a presentation as a Jordan open string
joining two different branes.  It is therefore interesting to confirm that all
roots of the $ADE$ algebras have Jordan open string presentations. We believe
this analysis improves substantially our understanding of the appearance of
the adjoint representations. As usual, selecting a base of simple roots 
involves choices. We make a choice and exhibit junctions  $\malpha_i$ 
representing the simple roots of the $ADE$ algebras, confirming that their
intersection numbers give rise to the appropriate Cartan matrices.

We define fundamental weight junctions $\momega^i$ dual to the simple root
junctions $\malpha_i$, and introduce additional junctions with nonzero
asymptotic charge to span the entire junction lattice. The new
junctions  can be thought of as additional fundamental weights $\momega^a$, and allow us
to develop an extended formalism.  When the metric in the lattice of junctions is
non-degenerate one can define dual roots $\malpha_a$, and the resulting extended Cartan
matrix is seen to be block diagonal.
An arbitrary junction can be expanded using the extended set of fundamental
weights; the coefficients of the $\momega^a$ give the asymptotic charges, 
and the coefficients of the $\momega^i$, we claim, are simply the Dynkin labels of the
associated weight vector. Proving this requires further analysis of the junction
lattice and its relation to the weight lattice.  Note that the claim implies that 
shifting a junction by an $\momega^a$ does not change its associated weight vector.

 We search systematically the lattices for all string junctions of
self-intersection $(-1)$, which we expect to be related to fundamental
representations.  
Any Jordan open string with one endpoint on a brane and the other at infinity
represents a junction of self-intersection $(-1)$.  Again, while it may not be
guaranteed, for all cases we have examined the junctions of self-intersection
$(-1)$ have Jordan open string representatives.

\smallskip
We then turn to the subject of obtaining general representations.
Even though we already have junctions $\momega^i$ for weights,
these are not proper junctions (except for $E_8$) as they require
basis strings with fractional coefficients.  Thus, given a weight
vector $\vec\lambda = \sum a_i \,\vec\omega^i$ one is naively led to 
a junction $\sum_i a_i \,\momega^i$ which is not  proper.
Here is where the junctions
$\momega^a$ carrying asymptotic charges  play a crucial role.  They themselves are not
proper, but can be added to the naive weight junction to give a proper junction representing
the particular weight.  More precisely, given a specific representation, one can find asymptotic
charges $(p,q)$ such that all weights in the representation are represented by proper
junctions.  In addition, we show that for a given algebra, every representation in each
conjugacy class can be represented by proper junctions using any one of a list of possible
$(p,q)$ values. The $(p,q)$ sets associated to each conjugacy class are disjoint and 
together they are complete. Moreover, they define an additive group;
given two sets, any sum of charges involving one element of each set
falls into a unique third set. 
Finally, the map associating conjugacy classes
to $(p,q)$ sets is an isomorphism between the group
of conjugacy classes under tensor product of representations, and the 
additive group of $(p,q)$ sets.  
Thus we elucidate the relation 
between the junction lattices and the corresponding Lie-algebra
weight lattice.

\medskip
The above understanding is obtained by explicit examination of particular
examples, followed by generalization. We aim to relate junction and weight lattices 
via a relation of the form $a_i = Q^\mu K_{\mu i}$, with $a_i$ the Dynkin labels defining
the weight vector associated to the junction with invariant charges $Q^\mu$.
The goal here is finding the matrix $K$. We give 
a detailed discussion for the case of the
$su(n)$ algebras and the case of the
$so(8)$ algebra.   The matrix $K$
 is partially fixed by the constraints of the 
adjoint representation. For the case of $su(n)$ the relation is completely
fixed by further consideration of the fundamental representation, whose states
we expect to be represented by asymptotic strings ending on the branes. In the 
case of $so(8)$ we see that the junctions of self-intersection $(-1)$
fit into the ${\bf 8_v}$, ${\bf 8_s}$ and ${\bf 8_c}$ representations in a 
unique fashion, and this fixes the matrix $K$ completely.

\smallskip
With these examples understood in detail we notice that 
in fact, $K$ has a simple interpretation: $K_{\mu i} = -(\ms_\mu , \malpha_i)$ 
is simply the intersection number between the basis string $\ms_\mu$ and the
junction $\malpha_i$ representing the root $\vec{\alpha_i}$.  
This allows a nice compact expression for the Dynkin labels of any
junction $\mJ$: one readily finds $a_i (\mJ) = - (\mJ , \malpha_i)$.  This
confirms the  prior claim  that for a given junction, the expansion coefficients of 
the $\momega^i$ are the Dynkin labels of the associated weight vector. This
expression for $K$ is uniquely fixed by the constraint that it gives the
correct assignment for the junctions belonging to the adjoint, and by the
constraint that two sets of junctions with different nonvanishing asymptotic charges 
are correctly matched to two representations. This
understanding enables us to deal quite easily with the algebras of the $E$
series.

\smallskip
We also consider the reverse problem, given some Dynkin labels specifying a
weight vector, how do we build the corresponding junction?  Here the key is
that we have two extra equations. Along with the Dynkin labels we think of the 
asymptotic charges $p$ and $q$ as two more labels.  The Dynkin labels of
the extended framework
then match the number of invariant charges. The equations can then be
inverted to give the invariant charges in terms of the extended Dynkin labels. 
At this stage one can see explicitly how suitable values of $p$ and $q$ are required to achieve
integral invariant charges and the connection to conjugacy classes of representations
becomes apparent.

\medskip
Before summarizing the organization of this paper we touch on a couple
of issues that were somewhat puzzling in \cite{GZ}. We wondered at that time
what was the precise Lie-algebraic counterpart of the combining of junctions. 
The answer is now
clear, combination of junctions is simply addition of the corresponding
lattice vectors,
and therefore equivalent to addition of the corresponding weight vectors. 
It is also clear that
while junctions associated to roots can always be combined, the result need
not be a root. Another point was the following: for the $E_{n+1}$ algebra the 
brane arrangement involved $n$ $A$-branes, one $B$-brane and two $C$-branes.
On the other hand the maximal subalgebra of $E_{n+1}$ of a similar type
is $su(n)\times u(1)\times su(2)$. This is a little surprising since
given the number and types of branes present one would have 
naively expected a $u(n)\times u(1) \times u(2)$ algebra. 
Thus out of three $u(1)$'s two are not realized.
Those $u(1)$ charges are two linear combinations of invariant charges under
which no $E_{n+1}$ root can be charged. It is clear that these must simply
be the $p$ and $q$ charges, expressed in terms of the invariant charges.
We wondered in \cite{GZ} how one could read the surviving
$u(1)$ charge from the picture of the junction. The answer follows from our
results, and for convenience we discuss it in an appendix. Given a junction,
we have already shown how to find its Dynkin labels for $E_{n+1}$, it then takes a
routine Lie-algebraic computation to derive the Dynkin
weights of subalgebras. In this way one can tell precisely which combination 
of invariant charges gives the $u(1)$ charge in the $su(n)\times u(1)\times su(2)$
subalgebra.

\smallskip
This paper is organized as follows. In section 2 we define invariant charges,
introduce the canonical presentation of junctions and the basis of strings.
In section 3 we introduce a symmetric bilinear intersection form and discuss
self-intersection numbers for junctions. After reviewing briefly some basics
of representation theory, we prove that the inequivalent junctions with zero
asymptotic charge and self-intersection $(-2)$ are in correspondance with
the roots of the enhanced symmetry algebra.  In section 4 we introduce
fundamental weight junctions and the extended formalism.  In section 5 we
search for all junctions of self-intersection $(-1)$. In section 6 we show 
how to relate Dynkin labels to invariant charges by discussing explicitly
$su(n)$, $so(8)$ and $so(2n)$. In section 7 we establish the relation between the
junction and weight lattices, and discuss the $E$ series. 
Finally in section 8 we offer some concluding remarks.

\section{Invariant charges and the lattice of junctions}\label{s:Invariant} 
\setcounter{equation}{0}

In this section we discuss the equivalence classes of junctions. 
Two junctions are considered equivalent if they can be related by {\it junction
transformations}. These transformations are of two types: moving pieces
of the junction across branes, and continuous deformations 
of the junction consistent with charge conservation. The
operation of brane crossing, the presently relevant
version of the Hanany-Witten effect~\cite{hanany}, can create or destroy 
string prongs \cite{GZ}. The continuous deformations of a junction include the
possibility of coalescing junction points (see also \cite{hauer}). 
Our strategy for the classification of junctions 
will be the following. We will recognize a set of invariants, the invariant
charges,  associated  to a junction. We will then
explain how any string junction can be given a {\it canonical presentation} 
determined uniquely by, and in one to one correspondance with the
invariant charges. 
This will prove that the invariant charges characterize completely the
equivalence classes of junctions.  
We will be led to define a basis of junctions comprised by strings, and
general junctions will be viewed as points in a lattice whose
generating vectors are the basis strings.

\subsection{Invariant charges}\label{ss:Invariant}

The invariants associated to a junction
will be a set of charges, one for each 7-brane in the background. 
Our intuition about charges is 
that a junction may be charged under the gauge field associated to a 
particular brane either if it has a prong on the brane, or if it crosses 
the branch cut associated to the brane. For each brane, the invariant
is simply the charge which combines the two possible 
contributions so that the charge is unchanged when 
the junction is
modified by crossing transformations. Such charges exist
because crossing transformations are indeed constrained
by charge conservation \cite{GZ}.

Consider a junction ${\bf J}$ and let $\mu$ index the various
branes in the configuration.
Let $[p_\mu, q_\mu]$ denote the labels of a brane, and let
${\cal C}_\mu$ denote the branch cut emerging from the brane and
oriented away from it. We now claim that the junction has
an invariant charge $Q^\mu(\mJ)$ associated to the $\mu$ brane given by
\begin{eqnarray}
Q^\mu ({\bf J}) &=& n_+ - n_-
 + \sum_{k=1}^{b_\mu} ( \,r_k \, q_\mu\,\,- s_k\, p_\mu\, )\,, \label{invch} \\
&=&  n_+ - n_-  +  \sum_{k=1}^{b_\mu} \left| {r_k\atop s_k}\,\,{p_\mu\atop q_\mu}
\right|\, . \nonumber
\end{eqnarray}
In the above expression, $n_+$ is the number of $({p_\mu\atop q_\mu})$ prongs
departing from the brane, and $n_-$ is the number of $({p_\mu\atop q_\mu})$
prongs ending on the brane. 
Moreover, $b_X$, an integer greater than or equal to zero, denotes 
the number of intersections of $\mJ$ with the branch cut ${\cal C}_X$.
Finally, $({r_k\atop s_k})$ is the charges of the
string belonging to $\mJ$ that crosses the cut at the $k$-th
intersection point, in a counterclockwise direction.

\begin{figure}
$$\BoxedEPSF{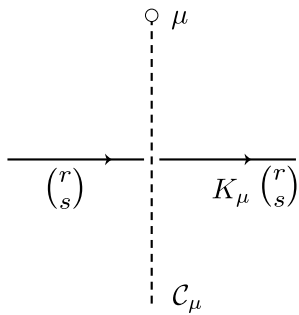 scaled 750}$$
\caption[Crossing rule]{The rule indicating how the charge of a string changes upon
crossing  counterclockwise the cut of a brane it cannot end on.}
\label{stringcrosscut}
\end{figure}

\smallskip
Let us now confirm this is the correct formula. The $(n_+ - n_-)$ term
in Eqn.~(\ref{invch}) gives the net number of outgoing prongs at the brane.
In addition, the $k$-th term in the sum equals
the number of outgoing
$({p_\mu\atop q_\mu})$ prongs that the $({r_k\atop s_k})$ string would generate
via the Hanany-Witten effect were it moved through the brane, so that it no
longer crosses the branch cut. 
As shown in Fig.~\ref{stringcrosscut}, upon 
crossing the branch cut ${\cal C}_\mu$ in the counterclockwise direction,
the $({r_k\atop s_k})$ string turns into
the $K_\mu({r_k\atop s_k})$ string where \cite{mgthbz} 
\begin{equation}
K_\mu = M_\mu^{-1} = 
\pmatrix{ 1+ p_\mu \, q_\mu & -p^2_\mu \cr q_\mu^2 & 1- p_\mu \, q_\mu}\,,
\label{hwcro}
\end{equation}  
and indeed 
\begin{equation}
K_\mu \pmatrix{r_k\cr s_k} =  \pmatrix{r_k\cr s_k} + 
( \,r_k \, q_\mu\,\,- s_k\, p_\mu\, ) \pmatrix{p_\mu\cr q_\mu}\,.
\label{confirmhw}
\end{equation}
This confirms that the $({r_k\atop s_k})$ piece of junction
can be moved across the
$\mu$ brane to a position where it does not intersect ${\cal C}_\mu$, by
adding $( \,r_k \, q_\mu\,\,- s_k\, p_\mu\, )\,$ {\it outgoing} prongs at $\mu$ that
reach the newly located piece of junction. Similarly, a string in the junction
that does not intersect ${\cal C}_\mu$ could be moved across the brane,
creating in the process a number of prongs. In can be readily verified
that in this case Eqn.~(\ref{invch}) would give two canceling contributions.
Note in particular that a crossing transformation may give rise to more 
than one prong.

\medskip
Equation (\ref{invch}) shows that the contributions from
the cut ${\cal C}_\mu$ only depend on the total $(p,q)$ charge being 
transported across it.
As we deform ${\cal C}_\mu$ continuously, the $(p,q)$ charge
transported across it will not change. This is clearly true even as we cross
junction points since charge is conserved at these points. 
We conclude that the $Q^\mu$ charge defined above is invariant under
deformations of the cut ${\cal C}_\mu$, as long as the deformation does
not require that the ${\cal C}_\mu$ cut cross another 7-brane. 

\subsection{Canonical presentation of junctions}\label{ss:Canonical}

\medskip
The above discussion of invariant charges 
suggests how to define a canonical presentation for any junction.
The idea is to transform any given junction  into an equivalent
junction that does not intersect any branch cut.
This is clearly possible since any specific intersection with a 
branch cut can be transformed away at the cost of introducing some prongs,
and this can be done with every intersection. 

\begin{figure}
$$\BoxedEPSF{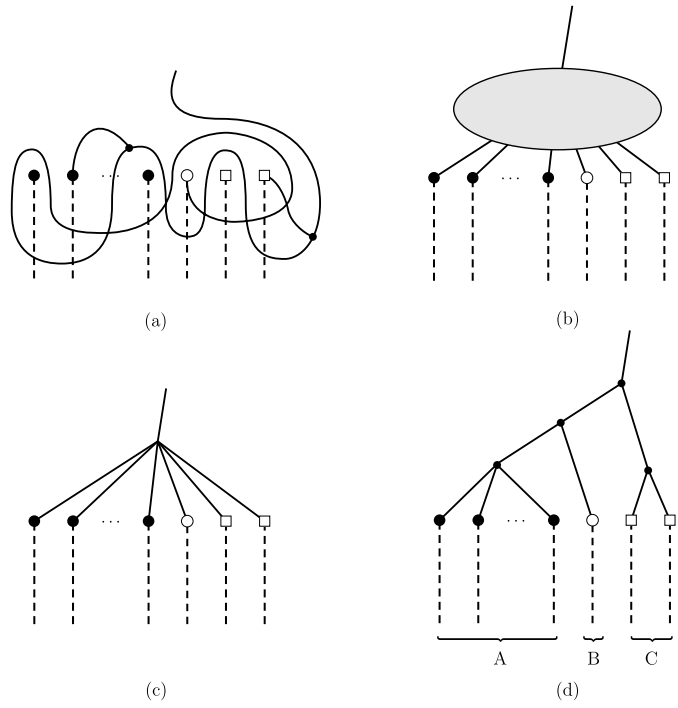 scaled 700}$$
\caption[Canonical presentation]{(a) A generic junction with an arbitrary
presentation. (b) Removing the crossing of branch cuts at the expense of prongs and
a possibly complicated graph shown shaded. (c) Shrinking the graph.
(d) Canonical presentation.}
\label{canonpres}
\end{figure}

Before doing this in detail we emphasize that we also present the 
branes in a canonical fashion. Following the conventions of \cite{GZ}
we use three types of branes, A-branes, B-branes and C-branes.
In our pictures, from left to right A-branes appear first, then B-branes
and finally C-branes. All branes have their cuts going downwards
vertically.  A-branes are represented by heavy dots, B-branes by
empty circles, and C-branes by empty squares (see Fig.~\ref{canonpres}).
Our conventions for monodromies are those of \cite{mgthbz} : 
\begin{eqnarray}
\label{expl}
{\bf A}=& [1,0]: \,\,K_A = & M_{1,0}^{-1} = T^{-1} = 
\pmatrix{1 & -1 \cr 0 & 1} \,, \nonumber\\ 
{\bf B}=& [1,-1]: K_B = & M_{1,-1}^{-1} =  S T^{2} 
             = \pmatrix{0 & -1 \cr 1 & 2} \,, \\
{\bf C}=& [1,1]: \,\, K_C = & M_{1,1}^{-1} =  T^{2}S 
             = \pmatrix{2 & -1 \cr 1 & 0} \,,\nonumber
\end{eqnarray}  
where $S$ is the matrix  
$$
S = \pmatrix{ 0 & - 1 \cr 1 & 0 } \,.
$$

We can now go back to the canonical presentation of  junctions.
Consider, for example, the
junction shown in Fig.~\ref{canonpres}(a).
After transformation
the junction  will look as in Fig.~\ref{canonpres}(b). At every brane
there is a set of prongs that converge towards a region where there could
be a complicated graph. Finally, there can be an outgoing string carrying 
he total charge of the configuration.  
Now we can allow the length of all the strings in the complicated
region to approach zero, reducing the graph to a single junction
point where all the prongs emerging from the various branes meet, as 
shown in Fig.~\ref{canonpres}(c).  
This configuration could be taken as the canonical presentation
but, for convenience, we will do a further step by introducing sub-junction
points.  All prongs of the leftmost set of mutually local branes will join
together forming a single string which is then joined by the string
resulting from  the prongs of the second set of mutually local branes, and 
so on. This is shown in Fig.~\ref{canonpres}(d).
We will call this the canonical presentation. By construction,
any junction is equivalent to its canonical presentation.   
As emphasized before, we do not address the issue of whether any
the junction has a BPS representative. Furthermore, if it has one, it
need not be the junction in the canonical presentation.

\smallskip
Note that the canonical presentation is determined {\it uniquely}
by the invariant charges that characterize the junction. In fact, the
invariant charge for any brane is precisely the number of outgoing prongs
in the canonical presentation. It follows that any two junctions
with the same invariant charges are equivalent because each is equivalent
to the same canonical presentation. Moreover, since the charges cannot
be changed by the allowed transformations, two junctions with different
charges are necessarily inequivalent. All in all, the invariant charges
give a complete classification of all possible equivalence classes
of junctions.

\subsection{The lattice of junctions}\label{ss:Thelattice}

\medskip
The canonical presentation suggests a natural way of building junctions
using a basis of {\it strings}. Let us label the branes using an index $\mu = 1 , \cdots
n$, where $n$ is the total number of branes in the configuration. For
a brane with label $\mu$ and of type
$[p_\mu, q_\mu]$,  we define the string $\ms_\mu$ as the outgoing $({p_\mu\atop
q_\mu})$ string starting at the brane and going off to infinity without crossing any
branch cut. This is a very special and simple junction, with invariant charges 
\begin{equation}
Q^\nu(\ms_\mu) = \, \delta^\nu_\mu  \,\,.
\label{basisst}
\end{equation} 
The string $(-\ms_\mu)$ is
the same string with reversed orientation.

\begin{figure}
$$\BoxedEPSF{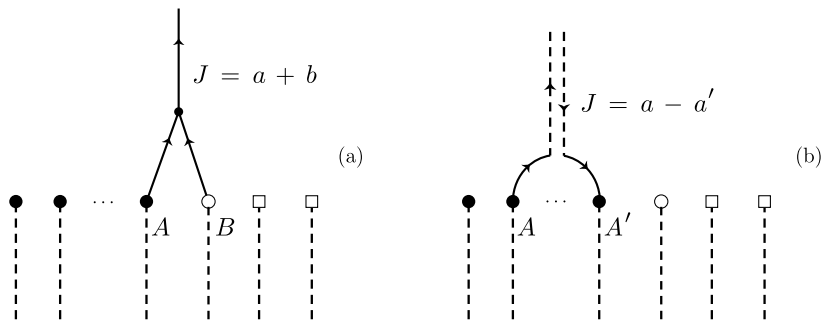 scaled 800}$$
\caption[Sample junctions]{(a) A junction obtained by adding an $\ma$
string to a
$\mb$ string. (b) The junction $(\ma - \ma')$ has no asymptotic charge and can be
thought as a string starting on the $A$ brane and ending on the $A'$ brane.}
\label{twojunctions}
\end{figure}

Let us define what is meant by adding junctions.  
 Given two junctions $\mJ_1$ and $\mJ_2$, we
say that $\mJ_1 + \mJ_2$ is the junction with charges 
\begin{equation}
 Q^\mu (\mJ_1 + \mJ_2) = 
Q^\mu (\mJ_1)+Q^\mu (\mJ_2)\,,
\label{addjunction}
\end{equation}
for all branes $\mu$ in the set of branes. In other words, we simply
add the prongs in the canonical presentation. This operation is
illustrated for two special examples in Fig.~\ref{twojunctions}. 
In the first example
we show the addition $(\ma + \mb)$ 
of two mutually nonlocal strings. Note that in the canonical presentation
this is a three string junction with outgoing prongs at the $A$ and $B$ branes
and one outgoing string carrying the total charge $({p_a + p_b\atop q_a+q_b})$.  
In the second example we consider $(\ma - \ma')$ where 
$A$ and $A'$ denote two branes of the same type. This time when we form
the canonical presentation for the sum the total outgoing charge vanishes.
This junction is simply a string departing from the $A$ brane and ending
on the $A'$ brane.

\smallskip 
It follows from these considerations that any junction $\mJ$ can
be written as a sum of basis strings, where the expansion coefficients
are the invariant charges.  We have   
\begin{equation}
\mJ = \sum_\mu \, Q^\mu(\mJ) \, \ms_\mu \,,
\label{expjunction}
\end{equation}
which is clearly in agreement with equation (\ref{addjunction}) by 
virtue of eq.~(\ref{basisst}). Since the charges $Q^\mu$ must be integers,
we have a {\it lattice of junctions}, with the set of basis strings serving
as basis vectors.
Each point in this lattice represents an equivalence class of junctions. 
Since each seven-brane of the background provides a basis string, the
dimensionality of the junction lattice equals the number of seven-branes.

As we proceed in our exploration of the lattice of junctions, we will
need objects having non-integral
expansion coefficients in the basis of strings.  These we shall call {\em
improper junctions}, to distinguish them from the proper junctions which
can correspond to physical states.

\section{A metric on the junction lattice and  Lie algebra roots}\label{s:metric}
\setcounter{equation}{0}

In this section we consider some additional structure
on the lattice of junctions.  We will introduce a symmetric
bilinear form $(\cdot \,, \cdot )$, {\it i.e.} an inner product, or a metric. 
We will define this inner product by giving all inner products of
basis strings.  This inner product will be sometimes referred to
as an intersection bilinear, since it originates from the intersection
bilinear on the two-dimensional submanifolds of $K3$ which are the F-theory
(or M-theory) lifts of the junctions. 
The discussion presented here focuses on the junctions 
and not on their lifts. 
We will call $(\mJ, \mJ)$ the self-intersection invariant of  the 
junction $\mJ$.  The relevance of the metric in the junction lattice is
that it is related to the metric in
weight lattice of the associated Lie algebra.  Some of this relation will be
explored in this section. A related discussion of self-intersection with a
different application in mind was 
given in \cite{nekrasov}. 

After a brief review of the relevant notions in Lie algebras, we show
how to identify the roots of the algebra for the various brane configurations.
Using the intersection bilinear
we can readily show that the inequivalent junctions of
self-intersection $(-2)$ are equal in number to the roots of the expected
Lie algebra. The junctions (or strings) corresponding to 
simple roots are easily identified in each case.

Since all roots are represented by junctions of self-intersection $(-2)$,
which is also the self-intersection number of any Jordan open string joining
two different branes, we are led to hypothesize that all roots of the $A,D,E$
series
of algebras have presentations as Jordan open strings. Many of these were known
previously \cite{johansen,GZ}, except for a class of roots in $E_7$ and
another class in
$E_8$. We construct explicitly Jordan open strings representing these roots,
thus confirming the hypothesis.  

\subsection{Intersection bilinear on the junction lattice}\label{ss:Intersection}

The intersection bilinear, defined on the equivalence classes of 
junctions, will be deduced by starting with an initial assignment 
for the self-intersection of basis strings, and then exploring the
implications of bilinearity and invariance under junction transformations.  
We declare the
self-intersection of any basis string $\ms$ to be minus one, 
\begin{equation}
\label{startp}
(\ms, \ms ) = -1 \,.
\end{equation}
To find out the value we ought to assign to the self-intersection
of a three string junction, we
can consider (as in the previous section) an $({r\atop s})$ string
crossing the cut of a $[p,q]$ brane. In this configuration
the string clearly has no self-intersection. If we move the string through
the brane, it no longer crosses the cut but we obtain prongs beginning on the
brane and ending on the string, forming some equivalent junction. 
Since we expect the self-intersection to be invariant under junction
transformations, the various contributions must cancel. The
contribution from the prongs is $[-(rq-sp)^2]$, where $(rq-sp)$ is the
number of prongs emerging from the brane
(see eq.~(\ref{confirmhw})).  To cancel this we must have a
contribution to the self-intersection arising from the junction point as follows. 

Consider the three string junction $\mJ_3$ defined as the joining of 
$({p_i\atop q_i})$ strings, with $i=1,2,3$, where all strings are ingoing
(or outgoing) and the label $i$ increases (cyclically) as we go around
the junction in the counterclockwise direction. 
Then the contribution to the self-intersection from the junction point
is given as 
\begin{equation}
\label{sijunc}
(\mJ_3, \mJ_3) = \left| {p_i\atop q_i}\,\,{p_{i+1}\atop q_{i+1}}
\right|\, .
\end{equation}
This result is independent of $i$ (with $p_4\equiv p_1$ and 
$q_4\equiv q_1$) and
can be readily verified to be invariant under $\sl2z$ transformations
of all the strings making up the junction.   The self-intersection of
a larger junction is the sum of the contributions from the various
three-string junctions constituting the large junction.  To this must be
added the contribution of the various prongs of the junction ending on branes.
One readily verifies that this prescription indeed gives a total self-intersection
of zero for the situation described in the previous paragraph.

This result is actually sufficient to deduce the intersection
matrix in the basis of strings. We consider the three standard types
of branes, $A, B$ and $C$, and corresponding basis strings 
$\ma_i, \mb_i$ and $\mc_i$. Consider now the junction 
$\mJ = \ma_i + \ma_j$, representing a basis string departing the
$A_i$ brane that joins at a junction point a basis string departing the 
$A_j$ string, with $j\not= i$. Its self-intersection has a contribution
of $(-2)$ from
the two prongs at the two branes, and zero contribution from the junction
point since the two strings joining there 
are of the same charge and the determinant
in (\ref{sijunc}) vanishes. Thus 
\begin{equation}
\label{deducemi}
-2 = (\ma_i + \ma_j \,, \ma_i + \ma_j)
= (\ma_i , \ma_i) + (\ma_j , \ma_j) + 2 (\ma_i , \ma_j)= -1-1
+ 2 (\ma_i , \ma_j) \,,
\end{equation}
showing that $(\ma_i, \ma_j )=0$ when $i\not= j$. In general we conclude
that 
\begin{equation}
(\ma_i, \ma_j ) = (\mb_i, \mb_j ) = (\mc_i, \mc_j ) = -\delta_{ij}\,.
\label{fsus}
\end{equation}
Consider now the junction $\mJ= \ma + \mb$ made by combining an arbitrary
$\ma$ string and an arbitrary $\mb$ string. It is important here to notice
that using the canonical presentation, at the junction point the counterclockwise
cyclic ordering of incoming strings is $\{ \ma, \mb, -\ma -\mb\}$. Thus
the self-intersection is
\begin{equation}
\label{deduceab}
(\mJ , \mJ ) = -1 -1 + \left| \matrix{1& 1\cr 0& -1} \right|\, = -3\,.
\end{equation}
On the other hand $(\mJ , \mJ ) = (\ma + \mb, \ma + \mb) = -2 + 2\, (\ma, \mb)$
and we therefore deduce that $(\ma , \mb) = -1/2$. Exactly analogous 
considerations give us
\begin{eqnarray}
(\ma \,, \mb ) &=& -1/2 \,, \nonumber \\
(\ma \,,  \mc) &=&  \,\,\,\, 1/2  \,, \label{ip} \\
(\mb \,, \mc ) &=&  \,\,\,\,\,\, 1 \, . \nonumber
\end{eqnarray} 
These relations define completely the intersection matrix for basis strings.
If $n_A, n_B$ and $n_C$ denote the number of $A, B$ and $C$ branes
respectively, we can define the set of basis strings $\{ \ms_\mu \}$ 
with $\mu = 1, 2, \cdots\, , n_A + n_B + n_C$, as
\begin{equation}
\label{matrixsi}
\{ \ms_\mu \}= \{ \ma_1, \cdots , \ma_{n_A}, \, \mb_1, \cdots , \mb_{n_B}, 
\mc_1, \cdots , \mc_{n_C} \} \, .
\end{equation}
With this notation, the intersection metric is given as a matrix
\begin{equation}
\label{imart}
S_{\mu\nu} = - ( \ms_\mu , \ms_\nu )\, = S_{\nu\mu} \,.
\end{equation}
For any pair of junctions $\mJ$ and $\mJ'$ we have 
\begin{equation}
\label{usethis} 
(\mJ , \mJ') = \sum_{\mu, \nu}  Q^\mu(\mJ) \, S_{\mu\nu} \, Q^\nu (\mJ') \, .
\end{equation}
For the case
of $so(8)$, where $n_A=4, n_B=n_C=1$, the metric is readily written with
the help of (\ref{ip}) and (\ref{matrixsi})  
\begin{equation}
\label{so8matrix}
S (so(8)) = \pmatrix{\hskip-3pt -1 & 0&0& 0& -1/2 &1/2 \cr
0 & -1&0& 0& -1/2 &1/2\cr
0 & 0&-1& 0& -1/2 &1/2\cr
0 & 0&0& -1& -1/2 &1/2\cr
\hskip-5pt-1/2 & -1/2& -1/2 & -1/2 & -1 &1\cr
1/2 &  1/2& 1/2 & 1/2 & 1 &-1}\,.
\end{equation}
As we can see, the last two rows just differ by a sign and thus  this
metric is degenerate. This will be the case as long as 
$n_B = n_C = 1$, and therefore the metric will be degenerate for brane configurations
leading to $so(2n)$ enhanced symmetry ($n_A = n)$.  For the  above matrix, 
in addition, the sum of the first four rows is proportional to the fifth one, and
thus the matrix has two
zero eigenvalues (and four negative eigenvalues, three of them equal
to minus one and one equal to minus three). 
As we shall see, there will be two linearly independent string junctions that
represent for $so(8)$  the two eigenvectors with eigenvalue zero. 
These junctions have
zero intersection number with any string junction. For $so(2n)$ with $n\geq 5$ there
is only one zero eigenvalue and one null junction.  

\smallskip
For the case of $E_6$ the metric is given by  
\begin{equation}
\label{e6matrix}
S (E_6) = \pmatrix{\hskip-3pt -1 & 0&0& 0&0& -1/2 &1/2&1/2 \cr
0 & -1&0& 0&0&  -1/2 &1/2& 1/2\cr
0 & 0&-1& 0&0&  -1/2 &1/2 &1/2\cr
0 & 0&0& -1&0&  -1/2 &1/2&1/2 \cr
0 & 0&0& 0&-1&  -1/2 &1/2&1/2 \cr
-1/2 & -1/2& -1/2 & -1/2& -1/2 & -1 &1&1\cr
1/2 &  1/2& 1/2 & 1/2& 1/2 & 1 &-1 &1\cr
1/2 &  1/2& 1/2 & 1/2& 1/2 & 1 &1 &-1}\,.
\end{equation}
This metric is nondegenerate; its determinant equals $+1/4$.
It does not have definite signature, however,  six  eigenvalues are negative
and two are positive. For the $E_{n}$ series, 
in general, one finds $n$ negative eigenvalues and two
positive eigenvalues, as will be confirmed later; the metrics for $E_7$
and $E_8$ are obvious generalizations of that for $E_6$.  

\medskip
We now give the expression for the self-intersection of an arbitrary junction
in terms of its invariant charges. We write
\begin{equation}
\mJ = \sum_{i=1}^{n_A} Q_A^i \, \ma_i  +  
Q_B \, \mb +  \sum_{i=1}^{2} Q_C^i \,\mc_i  \,,
\end{equation}
which is general enough for the present applications.  Using (\ref{ip})
we obtain 
\begin{eqnarray}
(\mJ , \mJ )  = && -  \sum (Q_A)^2 - (Q_B)^2  
- \sum (Q_C)^2   \nonumber\\
&& - Q_B \sum Q_A \, + (2 Q_B + \sum Q_A ) \sum Q_C \, . \label{jsii}
\end{eqnarray}
It is possible to simplify this expression by means of the total asymptotic
charge of the junction.  
Using $q = -Q_B + (Q_C^1 + Q_C^2)$ we find
\begin{equation}
\label{siformula}
(\mJ , \mJ ) = - \sum (Q_A)^2  \, + \, q \sum Q_A \, - q^2 \,+\, 2\, Q_C^1\,
Q_C^2 \,.
\end{equation}
This expression will enable us to show that the junctions of self-intersection
number $(-2)$ are in one-to-one correspondence with the roots of the expected
Lie algebras.

\begin{figure}
$$\BoxedEPSF{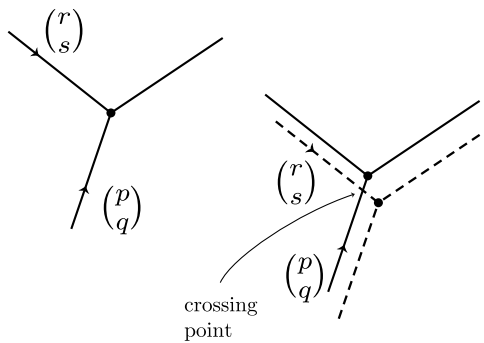 scaled 800}$$
\caption[Self-intersection]{Deriving the intersection number associated to crossing
strings by using the known self-intersection of a three-string junction.}
\label{splitjunctions}
\end{figure}

\medskip  
In computing the intersection number of two junctions directly from their
pictures, rather than from the general expression (\ref{usethis})
one must consider the
extra contributions that arise when strings simply cross without forming
a junction. One way to derive this contribution is
to reconsider the contribution of a junction point to the self-intersection,
and to interpret this contribution, as usual, as the intersection
of two slightly displaced versions of the junction (see Fig.~\ref{splitjunctions}). We
can see that the contribution from the junction point is equal to the contribution
due to the crossing of the $({p \atop q})$ and $({ r\atop s})$ strings.
We can make this unambiguous as follows.
At a crossing point we have two ingoing and two outgoing strings.
Pick either of these two groups and label the strings one and two, 
with string two following 
directly string one by counterclockwise rotation. 
The contribution to intersection is 
\begin{equation}
\label{crosscont}
\left| \matrix{p_1 & p_2 \cr q_1 & q_2 } \right| \, .
\end{equation}
To compute $(\mJ, \mJ')$ from the pictures of the $\mJ$ and $\mJ'$ junctions
we draw the junctions so that no junction points coincide, and we let the
asymptotic strings join at a junction point (an asymptotic junction) to form a single
asymptotic string. The intersection number has the
following  contributions: (i) $-nn'$ for each brane where $\mJ$ and $\mJ'$ have
$n$ and $n'$ outgoing prongs respectively, (ii) a contribution given by 
(\ref{crosscont}) for each crossing of strings, (iii) a contribution given
by{\it one half} of the result quoted in (\ref{sijunc}) for the asymptotic junction.
To compute the self-intersection $(\mJ, \mJ)$ from the picture of the junction $\mJ$
we add: (i) $-n^2$ for each brane where $\mJ$ has
$n$ prongs, (ii) a contribution given by 
(\ref{crosscont}) for each crossing of strings, (iii) a contribution given
by  (\ref{sijunc}) for each junction point.  We will not give a detailed derivation
of these rules here.

\subsection{Review of Lie algebras}\label{ss:Review}

In order to fix notation, we now give a brief review of some basic 
facts about simple Lie algebras.  For more background, see~\cite{lie}.
Consider a compact simple Lie algebra ${\cal G}$ of rank $r$ 
with Cartan generators
$\{H_i\}$, $i = 1 \ldots r$ satisfying
$[H_i, H_j] = 0$.
The remaining generators $\{E_\alpha\}$ satisfy 
$[H_i, E_\alpha] = \alpha^i\, E_\alpha$,  
and are thus characterized by the 
$r$ eigenvalues $\alpha^i$  defining the
{\it root vector} $\vec{\alpha}=  \alpha^i \,\vec{e}_i$ in an 
$r$-dimensional Euclidean space $E^r$. The finite set $\Phi$ of all roots is called the root system. 
The commutator $\left[ E_\alpha, E_\beta \right]$ can be nonzero
only if $\beta=-\alpha$, or if $\alpha + \beta \in \Phi$. 
The {\it base} $\Delta$ 
is a subset of $\Phi$ whose elements are $r$ 
{\it simple roots} forming a basis for $E^r$, with the
special property that all roots in $\Phi$ can be expanded in
coefficients that are either all negative or all positive.
We will be concerned henceforth only with the simply laced algebras
$A_n, D_n$ and $E_6, E_7$ and $E_8$, for which all roots have the same
length squared, conventionally chosen equal to two.

A state transforming under an arbitrary representation of 
${\cal G}$ is characterized by a weight vector 
$\vec{\lambda}= \lambda^i \, \vec{e}_i$, where $\lambda^i$ is the 
eigenvalue of $H_i$ on the state.  Each representation is characterized 
by a highest weight vector 
$\vec{\lambda_0}$;  all other weights 
are obtained from $\vec{\lambda_0}$ by subtracting  simple roots.
For a generic weight $\vec{\lambda}$ and a given simple root $\vec{\alpha}_i$,
there will exist numbers $p$ and $q$ such that $\vec{\lambda} + 
p \vec{\alpha}_i$ and $\vec{\lambda} - q \vec{\alpha}_i$ are also weights, 
but $\vec{\lambda}
+ (p+1) \vec{\alpha}_i$ and $\vec{\lambda} - (q+1) \vec{\alpha}_i$ are not.
Then one has $\vec{\lambda} \cdot \vec{\alpha}_i = q-p$.
The Cartan matrix for the algebras under consideration
is a symmetric matrix with integer entries defined by 
\begin{equation}
\label{cmatrix}
A_{ij} = \vec{\alpha}_i \cdot \vec{\alpha}_j\,,
\end{equation} 
whose information can be encoded 
in the Dynkin diagram, which for $ADE$ algebras is quite simple; nodes 
correspond to simple roots; nodes are  joined by a line if the 
corresponding roots are at an angle of
$120^\circ$, and are not joined if the roots are orthogonal.
Fundamental weights
$\vec{\omega}^i$ with $i=1, \cdots, r$ are defined as dual basis vectors
\begin{equation}
\label{fweights}
\vec{\omega}^i \cdot \vec{\alpha}_j=  \delta^i_j \,,
\end{equation}
and form the {\it Dynkin basis}.  The expansion 
of a weight vector in this basis reads
\begin{equation}
\label{dexpd}
\vec{\lambda} = \sum_i a_i\ \vec{\omega}^i\,,
\end{equation}
with integers $\{a_i\}$ called the {\it Dynkin labels}. Representations
are characterized by the Dynkin labels of the highest weight. Such labels
are necessarily non-negative. Note that, by construction
\begin{equation}
\label{dynkroots}
\vec{\alpha}_i = \sum_j A_{ij} \, \vec{\omega}^j\,,
\end{equation}
and the Cartan matrix is seen to have the Dynkin labels of the simple roots as
its rows.

\subsection{Roots and simple roots for the A${}_{\bf n}$ and 
D${}_{\bf n}$ algebras}\label{ss:Roots}

Jordan open strings beginning on one brane and ending on another one of the
same type manifestly have self-intersection
$(-2)$ and vanishing asymptotic charges.  
These strings represent  the
familiar roots of $A_n$ algebras.  
We will see that all roots, even those
in enhanced symmetry algebras, are represented by junctions
of self-intersection $(-2)$ and vanishing asymptotic charges. 
In this subsection we will search for all such junctions 
for the cases of $A_n$ and $D_n$ enhanced symmetry.
The simple roots are identified as a subset of the
junctions  whose intersection matrix is minus the Cartan matrix of the 
corresponding Lie algebra.

\subsubsection{The su(n) algebra}\label{sss:sun}

\begin{figure}
$$\BoxedEPSF{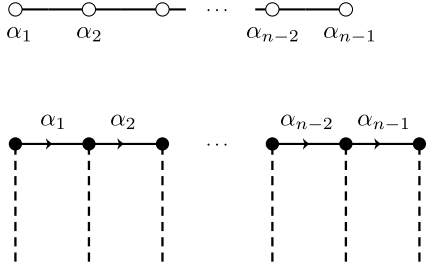 scaled 800}$$
\caption[Dynkin diagram and branes for $su(n)$]{Dynkin diagram for $su(n)$ and
associated brane configuration with junctions representing the simple roots.}
\label{dynkinsun}
\end{figure}

Consider first the case of $su(n)$, which is built from $n$ $A$ branes,
with no $B$ branes nor $C$ branes. In this case the self-intersection
formula (\ref{siformula}) reduces to 
\begin{equation}
\label{siso8}
(\mJ , \mJ ) = - \sum_{i=1}^n (Q_A^i)^2  = -2 \,.
\end{equation}
 In this case the vanishing asymptotic charge condition gives 
\begin{equation}
\label{subsisun}
0= p = \sum_{i=1}^n Q_A^i \, .
\end{equation}
The last two equations imply that the junctions must have two A prongs
one ingoing and one outgoing. Thus the complete list of solutions for
the inequivalent junctions is
\begin{equation}
\label{sunroots}
\pm ( \ma_i - \ma_j) \,,  \quad 1 \leq i < j \leq n \,,
\end{equation}  
making a total of $(n^2-n)$ junctions. As expected, this is precisely the number
of roots of $su(n)$. Note that they all have a simple interpretation as 
an open string extending between two different branes.
We can identify the simple roots as 
\begin{eqnarray}
\malpha_i = \ma_i - \ma_{i+1}\,, \quad i=1,\cdots , n-1.   \label{sunsimple}
\end{eqnarray}
This is in accord with the Cartan matrix of $su(n)$ since
$(\malpha_i , \malpha_{i+1}) = +1$.
The Dynkin diagram and the brane configuration with simple roots
are shown in Fig.~\ref{dynkinsun}.
In this case it is straightforward to see that all roots can be written
as linear combinations of simple roots with coefficients that are all
positive or all negative.  The highest root is 
\begin{equation}
\label{highestrootsu}
\sum_{i=1}^{n-1} \malpha_i   = \ma_1-\ma_n \,. 
 \quad (\mbox{\em highest
root})
\end{equation}  

\subsubsection{The so(2n) algebra}\label{sss:so2n}

\begin{figure}
$$\BoxedEPSF{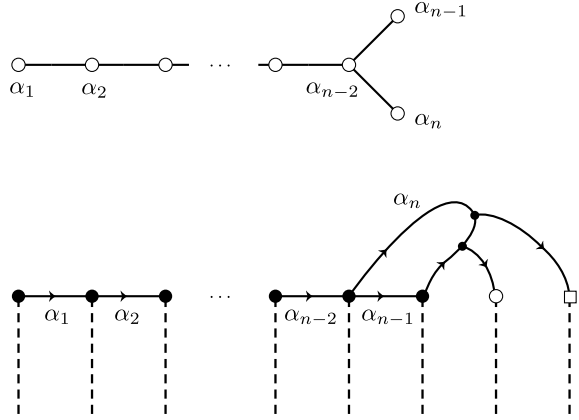 scaled 800}$$
\caption[Dynkin diagram and branes for $so(2n)$]{Dynkin diagram for $so(2n)$ and
associated brane configuration with junctions representing the simple roots.}
\label{dynkinso2n}
\end{figure}

Consider now the case of $so(2n)$ which contains $n$ branes of type $A$,
one brane of type $B$ and one brane of type $C$. 
In this case the conditions for roots are
\begin{equation}
\label{siso2n}
\sum_{i=1}^n (Q_A^i)^2 = 2  \,,\quad
\sum_{i=1}^n Q_A^i  + 2 Q_C = 0\,, \quad  Q_B = Q_C\, .
\end{equation}
Again, we must have two prongs on two different $A$ branes. If the two
prongs have opposite signs we must have $Q_B = Q_C =0$. These are
\begin{equation}
\label{sondirectroots}
\pm ( \ma_i - \ma_j) \,,  \quad 1 \leq i < j \leq n \,.
\end{equation}  
If the two prongs have the same sign, we need two more prongs of opposite
signs, one going to $B$ and one going to $C$
\begin{equation}
\label{sonindirectroots}
\pm ( \ma_i +  \ma_j - \mb -\mc ) \,,  \quad 1 \leq i < j \leq n \,.
\end{equation}  
The last two expressions give a total of $2(n^2-n)$ roots, the expected
number for $so(2n)$. It is easy to select a set of $n$ simple roots. We take
\begin{eqnarray}
\malpha_i &=& \ma_i - \ma_{i+1}\,, \quad i=1,\cdots , n-1 \,, \label{sonsimple} \\ 
\malpha_n &=& \ma_{n-1} + \ma_n  - \mb -\mc \nonumber\,.
\end{eqnarray}
It is clear that the intersection matrix of these roots generates the
Dynkin diagram of $so(2n)$.  In particular note that $\malpha_{n-2}$ 
is the root at the vertex;   
$(\malpha_{n-1},\malpha_{n}) =0$, and $(\malpha_{n-2},\malpha_{n}) = 1$,
as required. The Dynkin diagram and the brane configuration with simple roots
are shown in Fig.~\ref{dynkinso2n}. 
The simple roots generate all roots, but this time it
takes a little effort to see that the expansion coefficients are all
positive or all negative. In the above lists the roots with $(+)$ sign  in
front are positive roots, and the roots with $(-)$ sign in front are 
negative roots. The highest root in the adjoint representation
is
\begin{equation}
\label{highestroot}
\malpha_1 + 2 (\malpha_2 + \cdots + \malpha_{n-2}) + \malpha_{n-1} + \malpha_n
= \ma_1 + \ma_2 - \mb -\mc \,.  \quad (\mbox{\em highest root})
\end{equation} 

\subsection{Roots and simple roots for the E${}_{\bf n}$ algebras}\label{ss:Rooten}

This time we must consider the equations
\begin{eqnarray}
\sum_{i=1}^n (Q_A^i)^2 - 2 Q_C^1 \, Q_C^2 &=& 2  \,, \label{siexc1} \\
\sum_{i=1}^n Q_A^i  + 2 \sum_{j=1}^2 Q_C^j &=& 0\,, \label{siexc2} \\
Q_B-\sum_{j=1}^2 Q_C^j &=& 0 \,, \label{siexc3}
\end{eqnarray}
where the number $n$ of $A$ branes takes values $n=5,6,7$ for the cases of $E_6, E_7$ and
$E_8$ respectively.  
Note that a major simplification occurring in the previous subsection
does not take place here: the first equation does not imply immediately
a bound on the number of $A$ prongs allowed. We will now do a preliminary
analysis to show that indeed the above constraints do bound the number of
$A$ prongs.

Note first that (\ref{siexc2}) implies that the number of $A$ prongs 
must be even. Take  $\sum |Q_A| = 2k$ and consider the case when  
$k\geq 2$. It then follows from (\ref{siexc1}) that 
$Q_C^1 \, Q_C^2 >0$ and therefore $Q_C^1$ and $Q_C^2$ have the same
sign. From the (\ref{siexc2}) we find 
\begin{equation}
\label{ineq1}
\left| Q_C^1 + Q_C^2 \right| = \half \left| \sum Q_A \right| \leq  k\,.
\end{equation} 
Since $Q_C^1$ and $Q_C^2$ have the same
sign, we then have 
\begin{equation}
\label{ineq2}
Q_C^1 \, Q_C^2 \leq {k^2 \over 4} \,.
\end{equation}
On the other hand $\sum Q_A^2$ is bounded from below. We can go a long way
with a simple bound. We can take $Q_{A}^i = 2k/n$, for all $i$, to find 
$\sum_i Q_A^2 \geq 4k^2/n$. On account of (\ref{siexc1}) this gives 
\begin{equation}
\label{ineq3}
Q_C^1 \, Q_C^2 = {1\over 2} \sum_{i=1}^n (Q_A^i)^2 - 1 \,\geq \, {2k^2 \over n}-1 \,. 
\end{equation}
The last two inequalities can only be solved if
\begin{equation}
\label{ineq4}
k^2 \leq  {4n \over 8-n}\,.
\end{equation}
For $n=5$, the largest $k$ is $k=2$. This implies that for $E_6$ we
will at most have four $A$ prongs.  For $n=6$ we see that $k=2,3$ are
possible. Thus for $E_7$ we will have to consider the possibility of
up to six $A$ prongs. For $n=7$ it appears as if $k\leq 5$. Actually 
$k=5$ is not allowed.
Indeed, with $10$ prongs and 7 A-branes
$\sum Q_A^2 \geq 3 (2)^2 + 4 = 16$ (three
branes having two prongs and four branes each with a single prong). Then
eq.~(\ref{ineq3})  becomes
$Q_C^1 \, Q_C^2 \geq 7$ which is not compatible with eq.~(\ref{ineq2})
which now gives $Q_C^1 \, Q_C^2 \leq 25/4$. Therefore the maximal $k$ for
$E_8$ is $k=4$ corresponding to a maximum of eight prongs. This is
interesting since it requires that at least one $A$ brane have two or more
prongs.  

Note that $n=8$, which we would obtain by adding an
additional $A$ brane to the $E_8$ configuration, seems to allow all numbers
of $A$ prongs (see (\ref{ineq4})).
Certainly the number of $A$ prongs is not bounded, one can readily verify
that a total of $8k$ prongs, $k$ on each $A$ brane, is compatible with 
$Q_C^1 = -2k \pm 1$ and $Q_C^2 = -2k \mp 1$.  This phenomenon is likely
to  be related to the observations of Imamura \cite{imamura}.  

\subsubsection{The E${}_{\bf 6}$ algebra}\label{sss:e6}

\begin{figure}
$$\BoxedEPSF{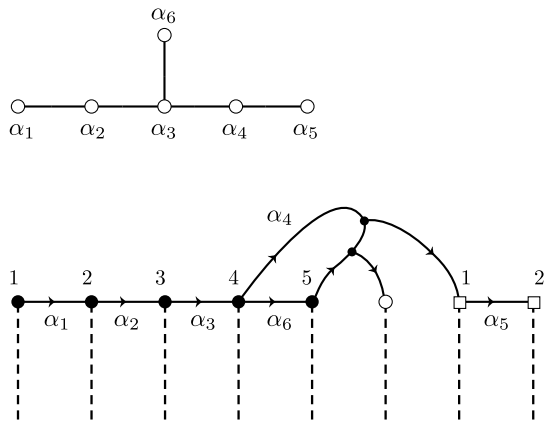 scaled 800}$$
\caption[Dynkin diagram and branes for $E_6$]{Dynkin diagram for $E_6$ and
associated brane configuration with junctions representing the simple roots.}
\label{dynkine6}
\end{figure}

We now consider each exceptional algebra in turn.  For $E_6$, the relevant
equations are
(\ref{siexc1}), (\ref{siexc2}), and (\ref{siexc3}) with $n=5$. 
In view of our previous analysis we must consider the case of zero, two 
or four $A$ prongs.  
Begin with the case of no $A$ prongs. We then 
need $Q_C^1 \, Q_C^2 = -1$, and we get two junctions
\begin{equation}
\label{e6roots}
\pm ( \mc_1 - \mc_2) \,.  
\end{equation}  
The case of two $A$ prongs is rather similar to the earlier one. Note that
the two prongs cannot be on the same $A$ brane for then $Q_C^1 \, Q_C^2 =1$,
and the second equation cannot be satisfied. The list of satisfactory
junctions with two $A$ prongs is 
\begin{eqnarray}
&& \pm ( \ma_i - \ma_j) \,,  \quad 1 \leq i < j \leq 5 \,. \nonumber \\
&&\pm ( \ma_i +  \ma_j - \mb -\mc_k ) \,,\quad 1 \leq i<j \leq 5 \,, k=1,2\,.
\end{eqnarray}  
This gives a total of $20 + 40 = 60$ junctions.
Finally, we must consider the possibility of having four $A$ prongs.
A short analysis of the first two constraint equations indicate that
there are no solutions unless the four
prongs fall on four different $A$ branes. It then follows from the first
equation that $Q_C^1 \, Q_C^2 = 1$ and the only integer solutions are
$Q_C^1 = Q_C^2 = \pm 1$ in which case $\sum Q_A = \mp 4$, by use
of the second equation. It thus follows that all $A$ prongs must be either
outgoing or ingoing. The junctions are then seen to be of the form
\begin{equation}
 \pm ( -\ma_i + \sum_{k=1}^5 \ma_k - 2\,\mb -\mc_1 -\mc_2 ) \,,  
\quad 1 \leq i  \leq 5 \,,  
\end{equation}  
for a total of 10 junctions. The complete solution set thus gives
$2+ 60 + 10 = 72$ which is indeed the number of roots in $E_6$.
We now choose a set of simple roots, 
\begin{eqnarray}
\malpha_1 &=& \ma_1 - \ma_2 \,, \nonumber \\
\malpha_2 &=& \ma_2 - \ma_3 \,, \nonumber\\
\malpha_3 &=& \ma_3 - \ma_4 \,, \nonumber\\
\malpha_4 &=& \ma_4 + \ma_5 - \mb -\mc_1 \,, \label{simpe6}\\
\malpha_5 &=& \mc_1 - \mc_2\,, \nonumber\\
\malpha_6 &=& \ma_4 - \ma_5\,, \nonumber
\end{eqnarray}
The corresponding string junctions are shown in Fig.~\ref{dynkine6}, and one
can readily verify that their intersection matrix precisely coincides
with the result encoded in the Dynkin diagram. 
It takes work to verify
that all the junctions listed above arise as positive or as negative 
linear combinations of the
simple roots, so we will not do this explicitly. We simply note that
the highest root is given by the following junction.
\begin{equation}
\label{highestroote6}
\malpha_1 + 2 \malpha_2 + 3\malpha_3 + 2\malpha_4 + \malpha_5 + 2\malpha_6
= \sum_{i=1}^4 \ma_i  - 2\, \mb -\mc_1 -\mc_2 \,.
 \quad (\mbox{\em highest root of} \, E_6)
\end{equation}
All roots of $E_6$ have familiar representations where their junctions
take the form of Jordan open strings between two branes.

\subsubsection{The E${}_{\bf 7}$ algebra}\label{sss:e7}

\begin{figure}
$$\BoxedEPSF{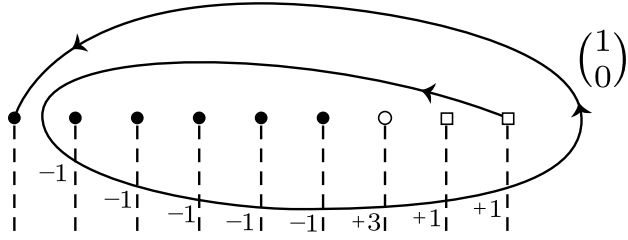}$$
\caption[Highest root of $E_7$]{A Jordan open string representing the highest root of
$E_7$ (see eq.~(\ref{highestroote7})). The
charge contributions from the cuts are indicated in the figure. This is a 
string beginning at a $C$ brane and ending on an $A$ brane.}
\label{jordan7}
\end{figure}

For $E_7$ we will now be briefer. A new kind of junction arises from 
(\ref{siexc1}), (\ref{siexc2}), and (\ref{siexc3}) with $n=6$. 
Again, a short computation
using the first two equations shows that no solution exists unless
each of the six prongs is on a different $A$ brane. In this case the 
first equation gives
$Q_C^1 \, Q_C^2 = 2$, which together with the second equation is 
solved by  $Q_C^1= \pm 2,  Q_C^2 = \pm 1$ or $Q_C^1= \pm 1,
 Q_C^2 = \pm 2$, 
together with $Q_A^i = \mp 1$ for all $i=1,\cdots 6$. 
The junctions are therefore
\begin{equation}
\label{frootse7}
\pm ( -\sum_{i=1}^6 \ma_i + 3 \mb + \sum_{i=1}^2 \mc_i  + \mc_k ) \,, \quad k=1,2\,.
\end{equation}   
It is of course straightforward to draw junctions representing the above 
states. In fact, a representation of the above states as a $BC$ open string was
given in \cite{GZ}. That string, however, had self-intersections. It is possible
to find a representative that is an open string without self-intersections. 
This is shown in Fig.~\ref{jordan7}, where we exhibit an $AC$ string with
appropriate charges. 

\begin{figure}
$$\BoxedEPSF{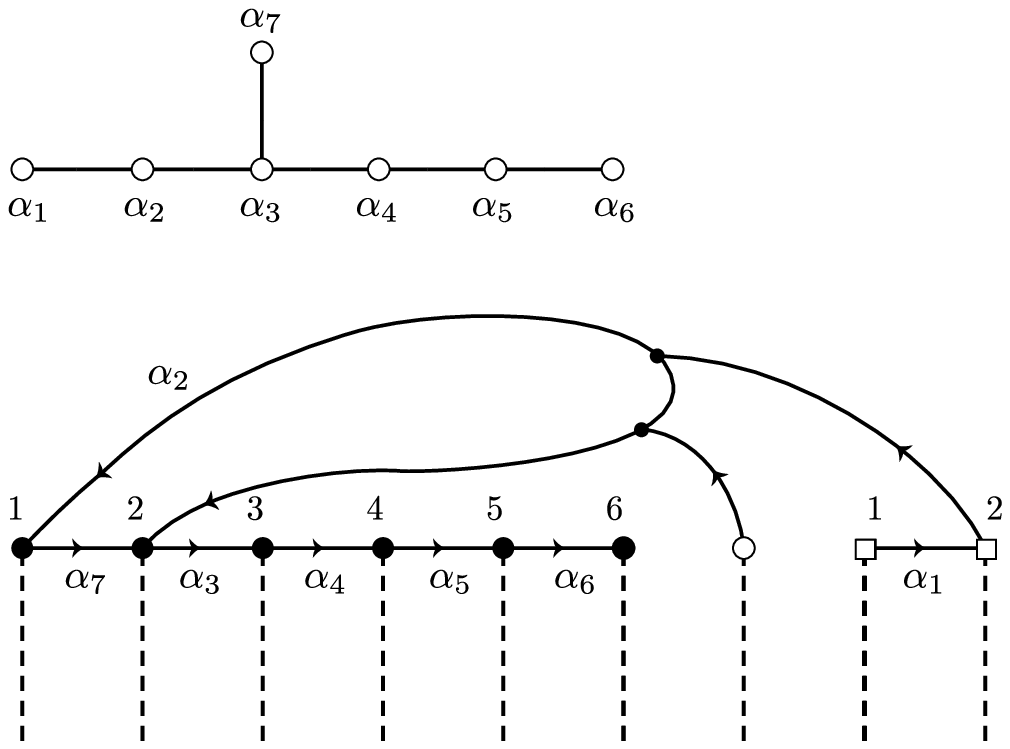 scaled 800}$$
\caption[Dynkin diagram and branes for $E_7$]{Dynkin diagram for $E_7$ and
associated brane configuration with junctions representing the simple roots.}
\label{dynkine7}
\end{figure}

The simple roots of
$E_7$ can be chosen as follows:
\begin{eqnarray}
\malpha_1 &=& \mc_1 - \mc_2 \,, \nonumber \\
\malpha_2 &=& -\ma_1 - \ma_2 + \mb + \mc_2 \,, \nonumber\\
\malpha_3 &=& \ma_2 - \ma_3 \,, \nonumber\\
\malpha_4 &=& \ma_3 - \ma_4 \,, \label{simpe7}\\
\malpha_5 &=& \ma_4 - \ma_5\,, \nonumber\\
\malpha_6 &=& \ma_5 - \ma_6\,, \nonumber\\
\malpha_7 &=& \ma_1 - \ma_2\, . \nonumber
\end{eqnarray}
These roots are shown in Fig.~\ref{dynkine7}, and one can see that their intersection
numbers generate the corresponding Dynkin diagram. The highest root of $E_7$ is 
\begin{eqnarray}
\label{highestroote7}
&&2\malpha_1 + 3 \malpha_2 + 4\malpha_3 + 3\malpha_4 + 2\malpha_5 + \malpha_6
+ 2\malpha_7 \nonumber\\
&&= -\sum_{i=1}^6 \ma_i + 2 \mc_1 + \mc_2 + 3 \mb \,.
 \quad (\mbox{\em highest root of} \, E_7) 
\end{eqnarray}

\subsubsection{The E${}_{\bf 8}$ algebra}\label{sss:e8}

\begin{figure}
$$\BoxedEPSF{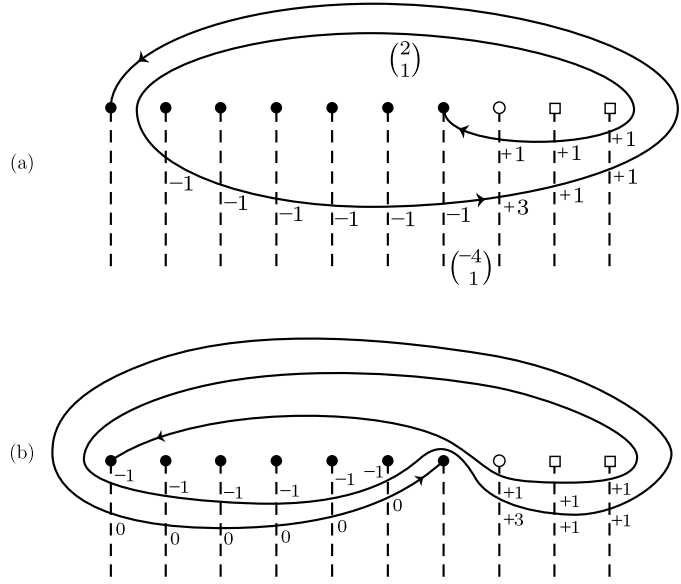 scaled 750}$$
\caption[Roots of $E_8$]{(a)  A Jordan $AA$ open string representing the state in
eq.~(\ref{frootse8}) with plus sign in front and $i=7$. This is actually
the highest weight of $E_8$.  
(b) A Jordan $AA$ open string for the state in eq.~(\ref{frootse8}) with plus sign in
front and $i=1$.}
\label{jordan8}
\end{figure}

For $E_8$ there is one interesting new possibility. As shown earlier, it
is possible to have eight $A$ prongs. Once more, straightforward analysis
of all possibilities shows that all eight prongs must have the same sign,
and that all branes have one prong, except for one that has two prongs.
Therefore $\sum Q_A = \pm 8$ and $\sum Q_A^2 = 10$. This requires
$Q_C^1 = Q_C^2 = \pm 2$ with $Q_A^i = \mp 1$ for six out of the
seven $A$ branes, and $Q_A = \mp 2$ for the other one.  The roots then 
read
\begin{equation}
\label{frootse8}
\pm ( -\sum_{k=1}^7 \ma_k - \ma_i  + 2 \mc_1 + 2\mc_2 + 4 \mb) \,, \quad
i= 1, \cdots 7\,.
\end{equation}  
It is possible to find open string representatives without self-intersections
for the above junctions. In Fig.~\ref{jordan8} we show two types of
strings.  The first type, illustrated in (a),  represents the state indicated
above with overall sign plus and $i=7$. A similar configuration can be 
used to represent the states with $i=2,3, \cdots , 6$. The state with $i=1$
requires a slightly different style of string, shown in part (b) of the figure.

\begin{figure}
$$\BoxedEPSF{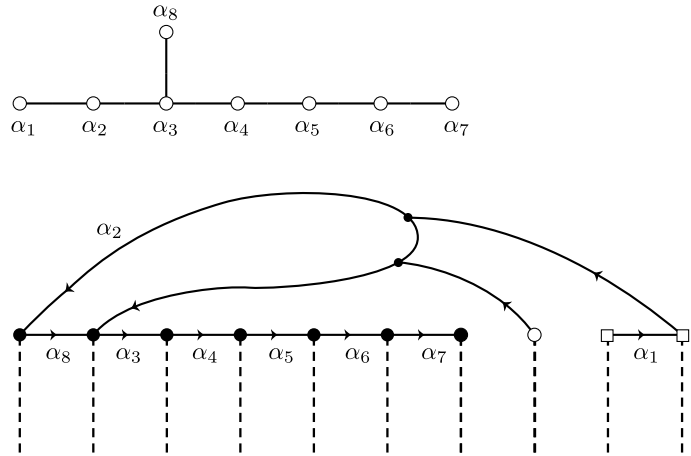 scaled 800}$$
\caption[Dynkin diagram and branes for $E_8$]{Dynkin diagram for $E_8$ and
associated brane configuration with junctions representing the simple roots.}
\label{dynkine8}
\end{figure}

The simple roots of $E_8$ can be chosen as follows:
\begin{eqnarray}
\malpha_1 &=& \mc_1 - \mc_2 \,, \nonumber \\
\malpha_2 &=& -\ma_1 - \ma_2 + \mb + \mc_2 \,, \nonumber\\
\malpha_3 &=& \ma_2 - \ma_3 \,, \nonumber\\
\malpha_4 &=& \ma_3 - \ma_4 \,, \label{simpe8}\\
\malpha_5 &=& \ma_4 - \ma_5\,, \nonumber\\
\malpha_6 &=& \ma_5 - \ma_6\,, \nonumber\\
\malpha_7 &=& \ma_6 - \ma_7\,, \nonumber\\
\malpha_8 &=& \ma_1 - \ma_2\,. \nonumber
\end{eqnarray}
These roots are shown in Fig.~\ref{dynkine8}, and one can see 
that their intersection
numbers generate the corresponding Dynkin diagram.  The highest weight vector
in the adjoint is given as:
\begin{eqnarray}
\label{highestroote8}
&&2\malpha_1 + 4 \malpha_2 + 6\malpha_3 
+ 5\malpha_4 + 4\malpha_5 + 3\malpha_6
+ 2\malpha_7 + 3\malpha_8 \nonumber \\
&&= -\sum_{i=1}^7 \ma_i  -  \ma_7 + 4\, \mb + 2\, \mc_1 + 2\, \mc_2 \,.
 \quad (\mbox{\em highest root of} \, E_8)
\end{eqnarray}

\section{An extended basis for the junction lattice}

The (proper) junctions representing roots do not form a complete basis
on the junction lattice.  Similarly, the associated dual junctions
representing fundamental weights, which are
not necessarily proper, do not form a complete basis. In the present section
we will introduce additional (not necessarily proper) junctions having
the interpretation of two new fundamental weights that together with the other 
weights make up an extended system of weights defining a complete basis of the
junction lattice\footnote{For the case of $A_n$ algebras only one extra weight 
is introduced.}. These additional junctions carry non-vanishing asymptotic
$(p,q)$ charges.  In fact we will see that the $p$ and $q$ labels play the
role of Dynkin labels for the two new junctions. The extended system of
weights has a well-defined intersection matrix, the analog of the inverse
Cartan matrix. This matrix does not always have an inverse. It does for the
case of the $A_n$ and $E_n$ algebras, but it does not for the case of the
$D_n$ algebras. When the matrix is nonsingular we can introduce two new 
junctions, also not necessarily proper, representing roots $\malpha_p$ and
$\malpha_q$ dual to the new weights.

In this extended formalism we will be able to conjecture the correspondance between
the junction and weight lattices. Moreover we will be able to relate the self-intersection
number of a junction to the length of the corresponding weight vector. 

\subsection{Junctions for asymptotic charges}\label{s:junctasymp}

In the previous section we identified junctions $\malpha_i$ for roots. 
These junctions carry zero asymptotic charges. 
Mimicking the definition of fundamental weights, we can use the inverse 
Cartan matrix to define
``junctions'' $\momega^i$ for fundamental weights as some linear combinations
of the root junctions.  These
will satisfy
\begin{equation}
\label{fwju} 
(\malpha_i \,, \, \momega^j ) = - \delta^j_i \,, 
\end{equation} 
along with
\begin{equation}
\label{bnb}
(\malpha_i \,, \, \malpha_j ) = - A_{ij} \,,\quad
 (\momega^i \,, \, \momega^j ) = - A^{ij} \,. 
\end{equation}
Note that the $\momega^i$ are not proper junctions since (except for
$E_8$) these are not integer linear combinations of basis strings; their
expansion coefficients in the string basis are rational, however.  
The $\momega^i$, being linear combinations of 
the  $\malpha_i$, carry zero asymptotic
charges and do not span the entire junction lattice. 
 
We are now going to introduce one special  junction  $\momega^p$ for each
$A_n$ algebra, and  two special junctions $(\momega^p, \momega^q)$ for
each algebra in the $D$ and $E$ series, to span the whole space.  These will
also generically be improper.  
We want these junctions to be orthogonal to the root junctions, 
\begin{eqnarray}
(\malpha_i , \, \momega^p ) &=& 0 \,, \nonumber\\ 
(\malpha_i , \, \momega^q ) &=& 0 \,. \label{njunn}
\end{eqnarray}
This orthogonality 
condition  suggests that these two junctions have the
interpretation of two new fundamental weights. 
Given that the weights $\momega^i$ are linear combinations of the roots
$\malpha_i$, we  also have    
\begin{eqnarray}
(\momega^i , \, \momega^p ) &=& 0 \,, \nonumber\\ 
(\momega^i , \, \momega^q ) &=& 0 \,. \label{opnjunn}
\end{eqnarray}

The $\momega^i$ together with
$(\momega^p, \momega^q)$ span the junction lattice. 
In order to fix the basis in the $(\momega^p, \momega^q)$ subspace 
we add one more constraint: we demand that
$\momega^p$  and  $\momega^q$ carry asymptotic charges $(1,0)$ and $(0,1)$
respectively, thus
\begin{eqnarray}
\label{cnjunn}
p\,(\momega^p ) = 1\,, &&  q\,(\momega^p ) = 0 \,,\nonumber   \\ 
p\,(\momega^q ) = 0\,, && q\,(\momega^q ) = 1   \,.   
\end{eqnarray}
As we will see, these conditions determine uniquely 
the junctions $\momega^p$  and  $\momega^q$.   
Since the inner product of fundamental weights defines the inverse
Cartan matrix, we define analogously
\begin{equation}
\label{necc}
 A^{ab} = \pmatrix{  A^{pp} &  
 A^{pq}\cr
 A^{qp} &  A^{qq}}  = 
 - \pmatrix{ (\momega^p ,
\momega^p) &   (\momega^p , \momega^q)\cr
(\momega^q , \momega^p) & (\momega^q , \momega^q)} \,.
\end{equation}
Having this matrix will be useful to relate the multiplicities of junctions
for different values of the asymptotic charges.

\subsection{Junctions in the extended formalism}

\smallskip 
Given a junction $\mJ$ it can always be expanded using the basis
$(\{ \momega^i \} , \momega^p , \momega^q )$. Since the weights
$\momega^i$ carry no asymptotic charge, such expansion will read
\begin{equation}
\label{jexp}
\mJ = \sum_i  a_i  \, \momega^i +  p\, \momega^p  + q\, \momega^q\, ,
\end{equation}
where $a_i$ are a set of numbers, $p= p(\mJ)$, and $q= q(\mJ)$.
We will show in sections 6 and 7  that   
constants $a_i$ are precisely the Dynkin
labels of the Lie algebra weight vector $\vec \lambda (\mJ)$
associated to the junction
$\mJ$; in other words we claim that the weight vector associated
to (\ref{jexp}) is
\begin{equation}
\label{wvass}
\vec \lambda (\mJ) =
\sum_i a_i \,\vec\omega^i\,.
\end{equation}
Included in this relation is our assertion that the $\momega^i$ are the
junctions corresponding to the fundamental weights, $\momega^i =
\mJ(\vec{\omega}^i)$.  Equations (\ref{jexp} ) and (\ref{wvass}) fix 
 completely the
map taking the junction lattice to the weight lattice.  
 It now follows
that the intersections of $\mJ$ with $\momega^p$ and $\momega^q$ 
only depend on the $p$ and $q$ charges of the junction.  We find
\begin{eqnarray}
\label{injup}
(\mJ , \momega^p) &=&  - p \, A^{pp} - q\,  A^{qp}\,, \nonumber\\
(\mJ , \momega^q) &=&  - p\,  A^{pq} - q\,  A^{qq}\,.   
\end{eqnarray}
Furthermore, we can easily calculate the intersection number of two junctions.
The self-intersection number of any junction follows from eqns.~(\ref{jexp}),
(\ref{bnb}) and  (\ref{necc}):  
\begin{equation}
\label{meqs}
- (\mJ , \mJ ) = \sum_{i,j} a_i \, A^{ij} \, a_j  +  p^2 A^{pp} + 2 pq\, A^{pq} + q^2\, A^{qq} \,,
\end{equation}
and on account of (\ref{wvass}) we get
\begin{equation}
\label{meq}
- (\mJ , \mJ ) = \vec\lambda (\mJ) \cdot  \vec\lambda (\mJ) +  p^2 A^{pp} + 2 pq\,
A^{pq} + q^2\, A^{qq}\,,
\end{equation}
relating the self-intersection of a junction to the length of the (claimed) associated
weight vector.

\subsection{Extended weights, extended roots, and extended Dynkin labels}

Having an extended system of weights, it is natural to  define an extended
inverse Cartan matrix $A^{\mu\nu}$. Construct the basis 
\begin{equation}
\label{eweights}
\{ \momega^\mu \}  =  \{  \{ \momega^i\} , \{ \momega^a  \} \}\,, 
\end{equation}
where the $\momega^a$ are $\{ \momega^p , \momega^q \}$, or just $\momega^p$
for the case of $A_n$. 
Then define
\begin{equation}
\label{jds}
 A^{\mu \nu} = - (\momega^\mu , \momega^\nu)\, .
\end{equation}
Note that in view of the orthogonality relations (\ref{opnjunn}) this extended
matrix takes the block diagonal form
\begin{equation}
\label{bdiag}
 A^{\mu\nu} = \pmatrix{ A^{ij} & 0 \cr 0 &  A^{ab} }\,.
\end{equation}
We will see that for the $D_n$ algebras the matrix $A^{ab}$ is
degenerate and we cannot define its inverse. On the other hand, for the
$E_n$ algebras $A^{ab}$ is nondegenerate and we can calculate its inverse.
We can then introduce roots $\malpha_a = \{ \malpha_p,\malpha_q\}$ dual to the
weights $\momega^a= \{ \momega^p, \momega^q\}$ such that
\begin{equation}
\label{dualbr}
(\malpha_a \,, \, \momega^b ) = - \delta^b_a\, .
\end{equation}
Since the $\malpha_a$ are linear combinations of the $\momega^a$
we have that
\begin{equation}
\label{orthn}
(\malpha_a , \malpha_i) = 0 \,, \quad (\malpha_a, \momega^i) = 0 \,.
\end{equation}
We have therefore obtained an extended set of roots
\begin{equation}
\label{eroots}
\{ \malpha_\mu \}  =  \{  \{ \malpha_i\} , \malpha_p , \malpha_q\}\,,
\end{equation}
dual to the extended weights
\begin{equation}
\label{edual}
(\malpha_\mu , \momega^\nu) = - \delta^\nu_\mu\,. 
\end{equation}
It now follows from (\ref{jexp})
and the last few equations that 
\begin{eqnarray}
\label{asymeqq}
p(\mJ) &=&   - (\mJ\,, \malpha_p)\,,  \\
q(\mJ) &=&  - (\mJ\,, \malpha_q) \,.
\end{eqnarray}
These two equations can be contrasted with the relation
\begin{equation}
\label{hui}
a_i (\mJ) = - (\mJ , \malpha_i)\,, 
\end{equation}
following from (\ref{jexp}) and  (\ref{edual}).  We are therefore
 led to define extended
``Dynkin'' labels
\begin{equation}
\label{edynkin}
\{ a_\mu \} \equiv  \{ \{ a_i\} , p , q\}\,,
\end{equation}
which enables us to rewrite  (\ref{jexp}) as
\begin{equation}
\label{ejunction}
\mJ = \sum_\mu \, a_\mu \, \momega^\mu\, , \quad a_\mu = - (\mJ , \malpha_\mu)\,.
\end{equation}
Note that this equation is simply a statement about 
junctions and their expansion in terms of fundamental weight junctions. 
We still have to show that the coefficients $a_i$ are indeed the Dynkin labels of
the weight vector associated to the junction $\mJ$.

\subsection{The case of su(n)}

In this case the only relevant asymptotic charge is $p$ and therefore
we introduce $\momega^p$, which is given by
\begin{equation}
\label{sjforsu}
\momega^p = {1\over n} \sum_{i=1}^n \ma_i \,,
\end{equation}
and satisfies both (\ref{njunn}) and (\ref{cnjunn}). For this weight
we find
\begin{equation}
\label{itsnn}
- (\momega^p \,, \, \momega^p ) = {1\over n} \,. 
\end{equation}
There is an associated root
\begin{equation}
\malpha_p = \sum_{i=1}^n \ma_i \,, 
\end{equation}
satisfying all the requisite properties.
 The extended inverse Cartan matrix and extended Cartan matrix  read 
\begin{equation}
A^{\mu\nu} = 
\pmatrix{A^{ij} & 0 \cr 0& 1/n} \,,\quad
A_{\mu\nu}  = \pmatrix{A_{ij} & 0 \cr 0& n}\,. 
\end{equation}
We can use this result to compute the intersection number of a junction. Using
(\ref{meq}) we find
\begin{equation}
-(\mJ \,, \mJ ) = \vec{\lambda}(\mJ) \cdot \vec{\lambda} (\mJ) + {1\over n}\,  p^2\,.
\label{relcon}
\end{equation}
Note that the right hand side is manifestly positive definite.

\subsection{The case of so(8)}

\begin{figure}
$$\BoxedEPSF{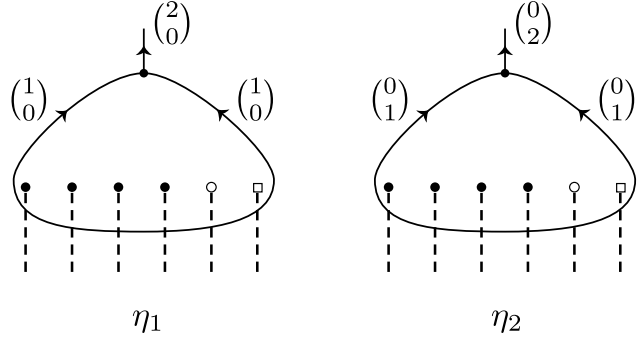 scaled 900}$$
\caption[Null junctions]{The two junctions $\mathbold{\eta}_1$ and
$\mathbold{\eta}_2$ having zero intersection with any junction. Their
self-intersection numbers are manifestly zero.}
\label{nulljunctions}
\end{figure}

We now turn to the case of $so(8)$, which we examine explicitly before
looking at the general case of $so(2n)$,
as its triality properties make it rather special.   Using the defining
equations (\ref{njunn}) and (\ref{cnjunn}), we find that the
fundamental weights read  
\begin{eqnarray}
\label{nullj}
2\,\momega^p \equiv \,\mathbold{\eta_1} &=& \mb + \mc \,, \\
2\,\momega^q \equiv \,\mathbold{\eta_2} &=& \sum_{i=1}^4 \ma_i - 3\, \mb - \mc \,.
\end{eqnarray}
We have introduced the above notation because we wish to emphasize
that $\mathbold{\eta_1}$ and $\mathbold{\eta_1}$ are proper junctions,
while $\momega^p$ and $\momega^q$ are not.  A simple computation shows
that $A^{ab}$ vanishes identically.  The junctions
 $\mathbold{\eta_1}$ and $\mathbold{\eta_1}$ are actually null junctions,
where by ``null'' we mean that their inner product with any junction vanishes.  
It is simple to verify that these are the unique null junctions by looking
for the most general junction with vanishing intersection with all basis
strings.  It is interesting to note that one can
have a presentation of $\mathbold{\eta}_1$ and $\mathbold{\eta}_2$
with manifest zero self-intersection. These are simply $({1\atop 0})$ and
$({0\atop 1})$
strings that go around all branes, cutting all branch cuts but having no
prongs.
The effect of the branch cuts is a monodromy $-{\bf 1}$ that is responsible
for giving
the total outgoing charges of $({2\atop 0})$ and $({0\atop 2})$ respectively.
This is shown in Fig.~\ref{nulljunctions}.

Given that the extended inverse Cartan matrix is degenerate we are
not able to introduce dual roots $\malpha_p$ and $\malpha_q$. This
is not a difficulty, it just means that the $(p,q)$ charges of a junction
cannot be obtained from a computation using the intersection bilinear.
This also means that the $\momega^p$ and $\momega^q$ junctions
do not contribute to the self-intersection, and we have that 
\begin{equation}
-(\mJ \,, \mJ ) = \vec{\lambda}(\mJ) \cdot \vec{\lambda} (\mJ)  \,.
\label{relconeso8}
\end{equation}
In $so(8)$ the self-intersection number of a junction equals the length
squared of the associated weight vector, regardless of the asymptotic charge.

\subsection{The case of so(2n) }\label{sss:caseso2n}

For the
case of
$so(2n)$ we  find extended fundamental weights
\begin{eqnarray}
\label{theonex}
2\, \momega^p &=& \mathbold{\eta}^1 = \mb + \mc \,, \nonumber\\
2\, \momega^q &=& \sum \ma_i - \mb + \mc - n\, \momega^p \,, 
\end{eqnarray}
where $\momega^p$ is null, but $\momega^q$ is not. In fact
$\momega^p$ is the unique null junction (up to normalization).
We can only introduce one new root, a root 
$\malpha_q$ satisfying $(\malpha_q, \momega^q) = -1$, and taking the form
\begin{equation}
\label{theone}
\malpha^q ={2\over 4-n} \Bigl( \sum \ma_i - \mb + \mc\Bigr)   \,, 
\end{equation}
and unique up to an irrelevant additive factor proportional to 
$\mathbold{\eta}^1$. Note that as expected, this root fails to exist for $n=4$.
It now follows by a simple computation that
\begin{equation}
-(\mJ \,, \mJ ) = \vec{\lambda}(\mJ) \cdot \vec{\lambda} (\mJ) 
- \fracs14 \, q^2 (n-4) \,.
\label{relconeso2n}
\end{equation}
This completes our discussion of the extended formalism
for $so(2n)$.

\subsection{The case of the E${}_{\bf n}$ algebras}\label{ss:extendeden}

For the $E_6, E_7$ and $E_8$ algebras the extended inverse Cartan matrix
is nondegenerate and one can introduce dual roots. The results for
these algebras are listed below. 

\subsubsection{The E${}_{\bf 6}$ algebra}\label{sss:casee6}

In this case we find 
\begin{eqnarray}
\momega^p &=& \hskip-6pt -{\fracs13}\,\sum_{i=1}^5 \ma_i + {\fracs43} \,\mb + {\fracs23} \, 
\sum_{i=1}^2 \mc_i, \label{w61}\\ 
\momega^q &=& \,\,\,\,\,\sum_{i=1}^5 \ma_i \,\,\,
 - 3\,\mb \,\,\, - \,\,\,\sum_{i=1}^2 \mc_i\,, 
\label{w62}
\end{eqnarray}
The $2\times 2$ block of the Cartan matrix is readily computed
\begin{equation}
\label{me6}
A^{ab} = \pmatrix{-1/3 & +1/2 \cr 1/2 & -1} \,,\quad
A_{ab} = \pmatrix{-12 & -6 \cr -6 & -4} \,. 
\end{equation}
We also find roots
\begin{eqnarray}
\malpha_p &=& -2\,\sum_{i=1}^5 \ma_i + 2 \,\mb -2 \, 
\sum_{i=1}^2 \mc_i, \label{en61}\\ 
\malpha_q &=& -2\,\sum_{i=1}^5 \ma_i + 4\,\mb \,.
\label{en62}
\end{eqnarray}
 Finally, the self-intersection of a junction is given by
\begin{equation}
-(\mJ \,, \mJ ) = \vec{\lambda}(\mJ) \cdot \vec{\lambda} (\mJ) - \fracs13 \,  p^2
+ pq - q^2 \,.
\label{relcone6}
\end{equation}

\subsubsection{The E${}_{\bf 7}$ algebra}\label{sss:casee7}

This time we find weights
\begin{eqnarray}
\momega^p &=& \hskip-6pt -{\fracs12}\,\sum_{i=1}^6 \ma_i + 2 \,\mb +  \, \,\,
\sum_{i=1}^2 \mc_i, \label{w71}\\ 
\momega^q &=&  {\fracs32}\sum_{i=1}^6 \ma_i \,
 - 5\,\mb \, - \, 2\,\sum_{i=1}^2 \mc_i\,.
\label{w72}
\end{eqnarray}
The corresponding $2\times 2$ intersection matrix is
\begin{equation}
\label{me7}
A^{ab} = \pmatrix{-1/2 & 1 \cr 1 & -5/2} \,,\quad
A_{ab} = \pmatrix{-10 & -4 \cr -4 & -2} \,, 
\end{equation}
and the associated roots read
\begin{eqnarray}
\malpha_p &=& -\,\sum_{i=1}^6 \ma_i  -2 \, 
\sum_{i=1}^2 \mc_i, \label{en71}\\ 
\malpha_q &=& -\,\sum_{i=1}^6 \ma_i + 2\,\mb \,.
\label{en72}
\end{eqnarray}
Finally, the self-intersection invariant can be computed as
\begin{equation}
-(\mJ \,, \mJ ) = \vec{\lambda}(\mJ) \cdot \vec{\lambda} (\mJ) - \fracs12 \,  p^2
+ 2\, pq - \fracs52 \, q^2 \,.
\label{relcone7}
\end{equation}

\subsubsection{The E${}_{\bf 8}$ algebra}\label{sss:casee8}

The extended weights are given by
\begin{eqnarray}
\momega^p &=& -\,\sum_{i=1}^7 \ma_i + 4 \,\mb +  \, 
2\, \sum_{i=1}^2 \mc_i, \label{w81}\\
\momega^q &=&  3\, \sum_{i=1}^7 \ma_i \,
 - 11\,\mb \, - \, 5\,\sum_{i=1}^2 \mc_i\,.
\label{w82}
\end{eqnarray}
The relevant Cartan matrix is 
\begin{equation}
\label{me8}
A^{ab} = \pmatrix{-1 & 5/2 \cr 5/2 & -7} \,,\quad
A_{ab} = \pmatrix{-28/3 & -10/3 \cr -10/3 & -4/3} \,. 
\end{equation}
The roots read
\begin{eqnarray}
\malpha_p &=& -\fracs23\,\sum_{i=1}^7 \ma_i  - \fracs23\,  \mb \, 
- 2\, \sum_{i=1}^2 \mc_i, \label{en881}\\ 
\malpha_q &=& -\fracs23 \,\sum_{i=1}^6 \ma_i + \fracs43\,\mb \,, 
\label{en82}
\end{eqnarray}
and contrary to the case of $E_6$ and $E_7$, are not proper junctions.
Finally, the self-intersection invariant can be computed as
\begin{equation}
-(\mJ \,, \mJ ) = \vec{\lambda}(\mJ) \cdot \vec{\lambda} (\mJ) -  \,  p^2
+ 5\, pq - 7 \, q^2 \,.
\label{relcone8}
\end{equation}

%We can now readily confirm that for $(p,q)= (1,0)$, there is only one 
%junction $\mJ$
%of self-intersection
%$(+1)$. Indeed, for $(p,q)= (1,0)$, the above equation gives
% $\vec{\lambda}(\mJ) \cdot
%\vec{\lambda} (\mJ) = 0$, and since the (standard) $E_8$ metric is positive definite
%there is only one solution,  $\vec{\lambda}(\mJ) = \vec 0$. We therefore have
%a unique junction and  all its associated Dynkin labels
%vanish. This completes the argument at the end of section 6.3. 

%%%%

\section{Junctions of self-intersection (--1)}\label{s:Junctions}

Having defined the extended formalism, we now turn to a search for
junctions with self-intersection $(-1)$.
A single string that begins on a brane and goes off to infinity without
self-crossings will have self-intersection $(-1)$, and if the brane it
ends on is one of several mutually local branes,  it will
transform in the fundamental of a manifest subalgebra.  
We would like to use self-intersection
$(-1)$ as a criterion for finding all the junctions that could be expected to
belong to the various fundamental representations of the enhanced Lie algebras.
We expect that all of these junctions will have presentations as simple Jordan
strings beginning on some brane and ending at infinity. Naturally, we no longer
constrain the asymptotic charges to vanish.

\subsection{Junctions of self-intersection (--1) for su(n)}\label{ss:Junctionsu}

Here the situation is rather trivial.  Self-intersection is just
\begin{eqnarray}
\left( \mJ, \mJ \right) &=& - \sum_{i=1}^{n} (Q_A^i)^2 \,,
\end{eqnarray}
and so $(\mJ, \mJ) = -1$ can only be solved by junctions with a single
$Q_A^i = \pm 1$, and all others vanishing.  These are just the junctions
in the fundamental and antifundamental representations, and the
asymptotic charge is simply $p = \pm 1$.

\subsection{Junctions of self-intersection (--1) for so(8)}\label{ss:Junctionso}

In principle we could look for junctions of self-intersection $(-1)$ for
any asymptotic charges $(p,q)$.  However, the null junctions $\meta_i$
found in (\ref{nullj}) make this unnecessary.  By definition they
have zero intersection with any junction; thus for any junction $\mJ$,
we can define the junction $\mJ'$,
\begin{equation}
\label{chla}
\mJ' = \mJ + m \, \mathbold{\eta}_1 + n \, \mathbold{\eta}_2\,,
\end{equation}
with $m,n \in \bbbz$, such that $(\mJ', \mJ') = (\mJ, \mJ)$.  Although
the self-intersection is unchanged under the transformation (\ref{chla}),
the asymptotic charges shift to $(p', q') = (p + 2m, q + 2n)$.  Thus
the junctions of a given self-intersection number are in one-to-one
correspondence for all $(p,q)$ values mod 2.
It is therefore sufficient to find the junctions for the $(p,q)$ values 
$(0,0)$, $(1,0)$, $(0,1)$, and $(1,1)$.

\medskip
For junctions of charges $(p,q) = (0,0)$ and self-intersection $(-1)$, we
need $\sum (Q_A)^2 =1$, together with
$\sum Q_A + 2Q_C =0$. The first equation implies that only one $Q_A$ is
nonvanishing, and thus the second equation cannot be solved.  There are
no junctions of appropriate self-intersection for this case.

\medskip

\begin{figure}
$$\BoxedEPSF{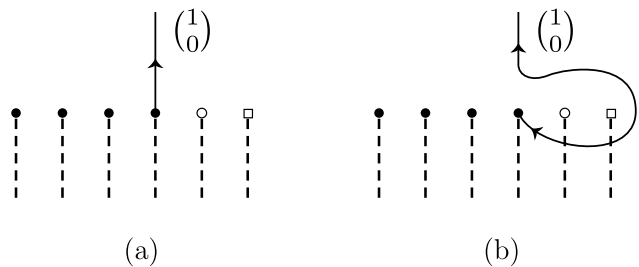 scaled 900}$$
\caption[Open strings for ${\bf 8_v}$]{String junctions with asymptotic charge
$(1,0)$. (a) The string $\ma_4$. (b)
The string $(-\ma_4 + \mb +\mc)$. These strings will be
shown to belong to the ${\bf 8_v}$.}
\label{string8v}
\end{figure}

For $(p,q) = (1,0)$ the conditions read
\begin{equation}
\label{sp1}
\sum (Q_A)^2 = 1 \,, \quad \sum Q_A + 2Q_C = 1\,, \quad Q_B = Q_C\, .
\end{equation}
We can have $Q_A = \pm 1$ for any one of the four A branes. This  gives
the following
eight junctions
\begin{equation}
\label{strs1}
\ma_i\,, \quad -\ma_i + \mb + \mc \,,  \quad i = 1, \cdots , 4\, ,   
\qquad\qquad ({\bf 8_v}) 
\end{equation}
where we have indicated in parenthesis the representation of $so(8)$ we 
will show these states belong to in section 6.2.  
It is possible to see explicitly that
these junctions have string representatives. We show two such
representatives in Fig.~\ref{string8v}.

\medskip

\begin{figure}
$$\BoxedEPSF{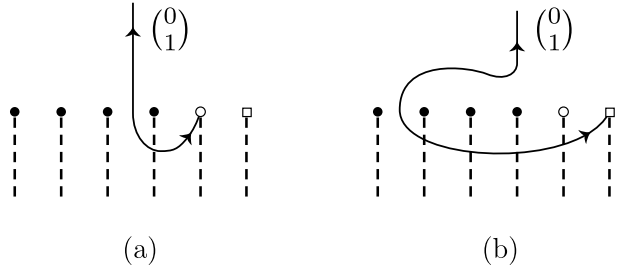 scaled 900}$$
\caption[Open strings for ${\bf 8_s}$]{String junctions with asymptotic charge
$(0,1)$. (a) The string $(\ma_4- \mb )$. (b)
The string $(\ma_2+\ma_3+\ma_4 -2 \mb -\mc)$. These strings will be
shown to belong to the ${\bf 8_s}$.}
\label{string8s}
\end{figure}

For $(p,q) = (0,1)$ the conditions read
\begin{equation}
\label{sp2}
\sum (Q_A)^2 = \sum Q_A \,, \quad \sum Q_A + Q_B + Q_C = 0\,, \quad -Q_B +
Q_C = 1\, .
\end{equation}
The first equation requires that there is at most one prong on any $A$
brane, and
that all prongs must be outgoing. The last two equations imply that
the number of $A$ prongs cannot be even. There are solutions for one or three
$A$ prongs, giving the following eight junctions
\begin{equation}
\label{strs2}
\ma_i - \mb\,, \quad  \sum_{k=1}^4 \ma_k - \ma_i - 2\,  \mb - \mc \,,
\quad i = 1, \cdots, 4\, . \qquad\qquad ({\bf 8_s}) 
\end{equation}
Once more,  it is possible to show explicitly that
these junctions have string representatives. Two 
representatives illustrating this are shown in Fig.~\ref{string8s}.

\medskip

\begin{figure}
$$\BoxedEPSF{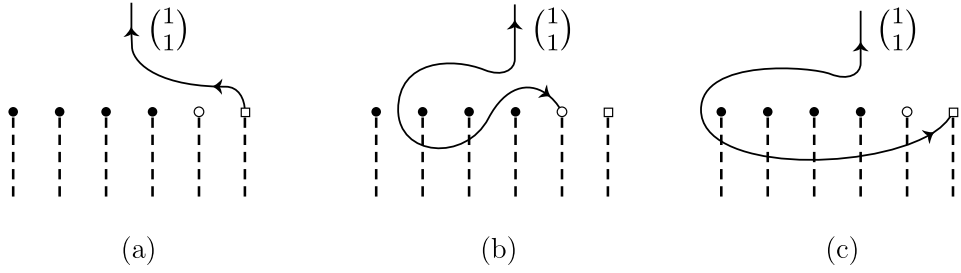 scaled 900}$$
\caption[Open strings for ${\bf 8_c}$]{String junctions with asymptotic charge
$(1,1)$. (a) The string $\mc$. (b)
The string $(\ma_2+\ma_3- \mb)$. (c) The string
$(\sum \ma_i - 2\mb -\mc)$. The strings in (a) and (c)
are singlets under $su(4)$.  The strings shown will be
seen to belong to the ${\bf 8_c}$.}
\label{string8c}
\end{figure}

Finally, for $(p,q) = (1,1)$ the conditions read
\begin{equation}
\label{sp3}
\sum (Q_A)^2 = \sum Q_A \,, \quad \sum Q_A + Q_B + Q_C = 1\,, \quad -Q_B +
Q_C = 1\,.
\end{equation}
This time
the number of $A$ prongs cannot be odd. If there are no $A$ prongs we
get one solution, if there are two $A$ prongs we get six solutions, and
if there are four $A$ prongs we get one solution. They read
\begin{equation}
\label{lone}
\mc \,, \quad \ma_i + \ma_j - \mb\,
\,\,( 1\leq i < j \leq 4)\, ,\quad \sum_{i=1}^4 \ma_i - 2\mb - \mc \,.\qquad
({\bf 8_c})
\end{equation}
Once more we have found a total of eight junctions.
Three string representatives are shown in Fig.~\ref{string8c}.

\subsection{Junctions of self-intersection (--1) for so(2n)}\label{ss:Junctionso2}

The algebra $so(2n)$ has a $2n$-dimensional vector representation and two 
$2^{n-1}$-dimensional spinors related by an outer automorphism of the algebra; 
only in the special
case $n=4$ do the dimensions of all three representations agree and the duality
automorphism is promoted to a triality.

For $n > 4$, there is just one null junction, given in (\ref{theonex}).
This allows us to shift the value of $p$ arbitrarily without changing a
junction's self-intersection.
One can shift the junction by an integer multiple of $\momega^q$,
\begin{eqnarray}
\mJ' = \mJ + m\ \momega^q \,,
\end{eqnarray}
and ask if self-intersection is preserved for specific values of $m$.  
We find that this requires $m = -2 q(\mJ)$, which
just flips the sign of the $q$ charge.  This is to be expected, as we can always
obtain a new junction with the same self-intersection by $\mJ \ra -\mJ$,
which sends $(p,q) \ra (-p,-q)$; $p$ can then be restored to its prior
value by shifting by $p\ \meta_1$.

It thus follows that to find all junctions of 
self-intersection $(-1)$ it suffices to analyze the cases $(0,q)$ and
$(1,q)$ for all nonnegative values of $q$.  
Here we will only investigate the cases $q=0,1$. 

The analysis proceeds exactly as 
for the special case of $so(8)$, eqns.~(\ref{sp1}) -- (\ref{lone}), with the
only difference being that there are now $n$ A-branes.  
Again, it is impossible to have junctions of self-intersection $(-1)$ and charges
$(0,0)$ for all $n$. For junctions
of charge $(1,0)$ there are $2n$ states that fit into the vector:
\begin{eqnarray}
\ma_i\,, \quad -\ma_i + \mb + \mc \,,  \quad i = 1, \cdots , n\, .   
\qquad\qquad ({\bf 2n})
\end{eqnarray}
For junctions with charges $(0,1)$ and $(1,1)$ we again find that there can
be no more than one prong on each A-brane, and each prong must be outgoing;
the number of A prongs must be odd and even, respectively.  As $n$ grows,
junctions with more and more prongs can be realized, resulting in 
representations whose dimensions go exponentially with $n$, exactly what
is required for the spinors: 
\begin{eqnarray}
\sum_{k=1}^{2k+1 \, \leq \, n} \, \left( n \atop 2k+1 \right) &=& 2^{n-1} \,, \\
\sum_{k=1}^{2k \, \leq \, n} \, \left( n \atop 2k \right) &=& 2^{n-1} \,,
\end{eqnarray}
making up the ${\bf 2^{n-1}}$ and ${\bf (2^{n-1})'}$ representations.
These junctions generalize those in eqns.\ (\ref{strs2}), (\ref{lone}),
having the form
\begin{eqnarray}
\ma_{i_1} + \ma_{i_2} + \cdots& + \ma_{i_m} - 
\frac{m+1}{2}\ \mb - \frac{m-1}{2}\ \mc  
\quad m\ {\rm odd} \,, \quad \quad &{\bf
2^{n-1}}\\
\ma_{i_1} + \ma_{i_2} + \cdots& + \ma_{i_m} - 
\frac{m}{2}\ \mb - \frac{m-2}{2}\ \mc \quad m\ {\rm even} \,, \quad \quad &{\bf
(2^{n-1})'}
\end{eqnarray}
with $i_1 < i_2 < \ldots < i_m$ and for all $m$ even or odd with $m \leq n$.

\subsection{Junctions of self-intersection (--1) for the exceptional
algebras}\label{ss:Junctione}

We now find the junctions of self-intersection $(-1)$
for the exceptional algebras. We will confirm
that these junctions appear in the right numbers to
allow us later on to identify them with fundamental representations (the ${\bf
27}$ and $\overline {\bf 27}$ of $E_6$, the  ${\bf 56}$ of $E_7$, and with a
predictable subtlety, the adjoint of $E_8$).  

Since the metrics on the corresponding junction lattices are non-degenerate,
there are no null junctions to consider.  Hence, in general, we cannot shift asymptotic
charges arbitrarily while preserving self-intersection.  Consider, anyhow,
shifting a junction as 
\begin{eqnarray}
\label{shiftJpq}
\mJ' = \mJ + r\ \momega^p + s\ \momega^q \,,
\end{eqnarray}
where $r$ and $s$ must
be chosen specially so that the resulting junction is proper.
Calculating the self-intersection of the new junction with the help of
(\ref{asymeqq}) we find 
\begin{eqnarray}
(\mJ', \mJ')  = (\mJ , \mJ) &-&2 \left(r\, p(\mJ)\, A^{pp} + s\, q(\mJ)\,
A^{qq}\right) \\ &-& 2 \left( s\, p(\mJ)\, + r\, q(\mJ)\,   \right) A^{pq} \\
&-& r^2\, A^{pp} - s^2\, A^{qq} - 2 r\, s\, A^{pq} \,.
\nonumber
\end{eqnarray}
We will be particularly interested in the case when junctions related by
(\ref{shiftJpq}) have the same self-intersection number.  Given a junction
$\mJ$ of fixed $(p,q)$ charges, this will only happen, if at all, for some
specific values of $r$ and $s$.

It is clear that one can rewrite (\ref{shiftJpq}) to view $\mJ$
as the result of a transformation on $\mJ'$, with integers $r'=-r$ and
$s'=-r$.  As a result, if one has determined exhaustively the junctions
$\{\mJ\}$ of a given self-intersection $u$ with some charges $(p,q)$, and one
finds a transformation mapping these to junctions $\{\mJ'\}$ with
self-intersection $u'$ and charges $(p',q')$, one can be sure that there are no
other junctions with self-intersection $u'$ and charges $(p',q')$; if there
were, one could turn the transformation around and find new junctions
with charges $(p,q)$ and self-intersection $u$, contradicting the fact that
these were all known.

We will not explicitly examine all possible asymptotic charges.  We consider
several cases when $p\not= 0$ and $q=0$, and then afterwards relate these
to other asymptotic charges using transformations as above.  For these
asymptotic charges the equations are 
\begin{eqnarray}
\sum_{i=1}^n (Q_A^i)^2 - 2 Q_C^1 \, Q_C^2 &=& 1  \,, \\ 
\sum_{i=1}^n Q_A^i  + Q_B +\sum_{j=1}^2 Q_C^j &=& p \,, \\
Q_B-\sum_{j=1}^2 Q_C^j &=& 0 \,.
\end{eqnarray}
The first equation is the self-intersection, and the other two are the
constraints on the charges.  Here $n=5,6,7$ for $E_6$, $E_7$, $E_8$ respectively.

If $(p,q) = (0,0)$, we obtain $\sum_{i=1}^n (Q_A^i)^2 - 2 Q_C^1 \, Q_C^2 = 1$, 
which requires $\sum_{i=1}^n (Q_A^i)^2$ odd, as well as $\sum_{i=1}^n Q_A^i +
2\sum_{i=1}^2 Q_C^j = 0$, requiring $\sum_{i=1}^n Q_A^i$ to be even.
One can readily check that satisfying both conditions is impossible.  Thus
there are no junctions of self-intersection $(-1)$ with charges $(0,0)$.

Now examine $(p,q) = (1,0)$.  The equations are
\begin{eqnarray}
\sum_{i=1}^n (Q_A^i)^2 - 2 Q_C^1 \, Q_C^2 &=& 1  \,, \label{eq1} \\
\sum_{i=1}^n Q_A^i + 2\sum_{j=1}^2 Q_C^j &=& 1 \,, \label{eq2}  
\end{eqnarray}
with $Q_B = \sum Q_C$.  Both $\sum_{i=1}^n (Q_A^i)^2$ and $\sum_{i=1}^n Q_A^i$
must be odd, which is possible.  Define the total number of prongs
$\sum_{i=1}^n | Q_A^i | = 2k + 1$.  Then 
\begin{eqnarray}
- (2k + 1)  \leq \sum_{i=1}^n Q_A^i \leq 2k + 1 \,,
\end{eqnarray}
or, using~(\ref{eq2})
\begin{eqnarray}
k + 1  \geq \sum_{j=1}^2 Q_C^i \geq -k \,.
\end{eqnarray}
Then we square and use the triangle inequality to obtain
\begin{eqnarray}
Q_C^1 \, Q_C^2 \leq \frac{(k+1)^2}{4} \,.
\end{eqnarray}
Now the lower bound on $\sum (Q_A)^2$ is obtained by setting $|Q_A^i| = (2k+1)/n$ for all $i$.  Then
\begin{eqnarray}
Q_C^1 \, Q_C^2 \geq \half \left(\frac{(2k+1)^2}{n} - 1 \right) \,.
\end{eqnarray}
The only way to satisfy both inequalities is
\begin{eqnarray}
(n-8) k^2 + (2n-8) k + 3n -2 \geq 0 \,.
\end{eqnarray}
For the three exceptional algebras we thus have
\begin{eqnarray}
\label{nprongs}
E_6:  &k = 0,1,2 \ra & 1,3,5\ {\rm prongs}, \nonumber \\ 
E_7:  &k = 0,1,2,3,4 \ra & 1,3,5,7,9\ {\rm prongs}, \\ 
E_8:  &k = 0,1,2,3,4,5,6,7,8 \ra & 1,3,5,7,9,11,13,15,17\ {\rm prongs}. \nonumber 
\end{eqnarray}
We still must satisfy the self-intersection equation, however, and not
every number of prongs permitted by the inequalities can be realized
on the brane configuration.  In Table 1 we list the possible patterns of 
junctions,
with the number of distinct junctions for each algebra listed at right.
At the bottom we find the total number of junctions with self-intersection
$(-1)$ for each.

\begin{table}
\begin{center}
\def\st{\vrule height3ex depth 2ex width0pt}
\begin{tabular}{|c|c|c|c|c|} \hline
Junction $\mJ$ with $(\mJ ,\mJ) = -1$ & Conditions & $E_6$ & $E_7$ & $E_8$ \st\\ \hline \hline
$\ma_i$ & $1 \leq i \leq n$ & 5 & 6 & 7 \st\\ \hline
$-\ma_i + \mb + \mc_j$ &  $1 \leq i \leq n$,~ $1 \leq j \leq 2$ & 10 & 12 & 14 \st\\
\hline
$-\sum_{p=1}^3 \ma_{i_p}  +2\mb + \mc_1 + \mc_2$ 
& $1 \leq i_1 < i_2 < i_3 \leq n$
& 10 & 20 & 35 \st\st\\ \hline
$-\sum_{p=1}^5 \ma_{i_p} 
+3\mb  + \mc_k + \sum_{j=1}^2 \mc_j$  & 
\begin{tabular}{c} $1 \leq i_1 < \cdots < i_5 \leq n$, \st\\%%%%
$1\leq k \leq 2$ \st\end{tabular} & 2 & 12 & 42 \st\\ 
\hline
$ -\ma_i -\sum_{p=1}^6 \ma_{i_p}  +4\mb + 2\sum_{k=1}^2 \mc_k$ &
\begin{tabular}{c}
  $1 \leq i_1 < \cdots < i_6 \leq n$, \st\\%%%%
$i = i_p\, (p=1, \ldots , 6) $ 
\st\end{tabular}
 & --- & 6 & 42 \st\\ \hline 
$-\sum_{i=1}^7 \ma_i + 4 \mb  + 2 \mc_k + \sum_{j=1}^2 \mc_j$ 
& $1 \leq k \leq 2$ & --- & --- & 2 \st\\ 
\hline
\begin{tabular}{c} 
$- \ma_{i_1} - \ma_{i_2} -\sum_{i=1}^7 \ma_i  $ \st\\%%%%
$ + 5 \mb + \mc_l + 2 \sum_{k=1}^2 \mc_k $ \st
\st\end{tabular} 
& $1 \leq i_1 < i_2 \leq 7$,~ $1 \leq l \leq 2$ & --- & --- & 42 \st\\ 
\hline
\begin{tabular}{c} 
$-2\sum_{i=1}^7 \ma_i + \sum_{p=1}^3 \ma_{i_p} $ \st\\%%%%
$ + 6 \mb + 3 \sum_{k=1}^2 \mc_k$ \st
\st\end{tabular} & $1 \leq
i_1 < i_2 < i_3 \leq 7$ & --- & --- & 35 \st\\ 
\hline
\begin{tabular}{c} 
$ \ma_i -2\sum_{k=1}^7 \ma_k  + 7 \mb$ \st\\ %%%%
$ + \mc_l + 3 \sum_{k=1}^2 \mc_k$ \st
\st\end{tabular} 
& $1 \leq i \leq 7$,~ $1 \leq l
\leq 2$ & --- & --- & 14 \st\\  
\hline
\begin{tabular}{c} 
$ - \ma_i -2\sum_{k=1}^7 \ma_k + 8 \mb$ \st\\%%%%
$+ 4 \sum_{k=1}^2 \mc_k$ \st
\st\end{tabular} 
& $1 \leq i \leq 7$ & --- & --- & 7 \st\\
\hline 
\hline 
Total junctions & & 27 & 56 & 240 \st\\ 
\hline
\end{tabular}
\end{center} 
\label{entable}
\caption{Junctions of self-intersection $(-1)$ and charges $(1,0)$ for the 
exceptional algebras.}
\end{table}

For $E_6$ the number of  junctions can precisely fill out a ${\bf 27}$.
In the case of $E_7$, we have a number of junctions that can fill the
${\bf 56}$.  Note
that although the bounds above suggested that nine A-prongs was possible
for the case of $E_7$, there is no
realization on just six A-branes.

Finally, for $E_8$, we have a total of 240 junctions,
and configurations with 17 A-prongs cannot be realized.
The smallest representation of
$E_8$ is the adjoint, the ${\bf 248}$.  We are missing eight junctions to 
complete this representation.  We will see in section \ref{ss:Dynkine8} 
that these eight junctions
correspond to the Cartan generators.  They cannot be realized as junctions
of self-intersection $(-1)$; in fact we will see that for the present value of
the
asymptotic charge they are realized as a unique junction 
(appearing with multiplicity eight) of self-intersection $(+1)$. 
This is perhaps not too
surprising when we recall that while junctions representing roots have
self-intersection $(-2)$, Cartan generators are  junctions of zero self-intersection.

We have considered only asymptotic charges $(0,0)$ and $(1,0)$.  As we will
see in more detail later, junctions with a given asymptotic charge
can only produce representations of a certain conjugacy class.  $E_8$ has
just one conjugacy class and $E_7$ has two, but for $E_6$ we have three
and we would like to find a representation in the third conjugacy class, the
class containing the ${\bf \overline{27}}$ .
We obtain junctions of self-intersection $(-1)$
for this representation  by just changing the sign of
every invariant charge in the junctions for the ${\bf {27}}$; in
this case the asymptotic charge is $(-1,0)$.   

We can obtain junctions of other asymptotic charges from the transformations
(\ref{shiftJpq}).  For $E_6$ with initial charges $(1,0)$, we find 
\begin{eqnarray}
\label{e6shift}
(\mJ^{(1 + r,s)},\mJ^{(1 + r,s)}) = (\mJ^{(1,0)},\mJ^{(1,0)}) +\fracs13 r^2 + s^2  - r\, s +
\fracs23 r - s \,.
\end{eqnarray}
Recall that (\ref{w61}) requires $r=0$ (mod 3) to obtain proper junctions. 
One solution is $r=0, s=1$ which produces 27 junctions with $(-1)$
self-intersection and $(1,1)$ asymptotic charge.  As we explained before,
these exhaust the junctions at the given self-intersection and asymptotic
charge.

For $E_7$ with the same initial charges, we arrive at
\begin{eqnarray}
\label{e7shift}
(\mJ^{(1 + r,s)},\mJ^{(1 + r,s)}) = (\mJ^{(1,0)},\mJ^{(1,0)}) + \fracs12 r^2 +
\fracs52 s^2 - 2r\, s + r - 2s \,.
\end{eqnarray}
(\ref{w71}), (\ref{w72}) require that $r + s = 0$ (mod 2) for proper
junctions.  A possibility is $r=1, s=1$, resulting in 56 junctions of
asymptotic charges $(2,1)$.

Finally we consider $E_8$.  Instead of $(1,0)$, begin with $(0,0)$.  We find
\begin{eqnarray}
\label{e8shift}
(\mJ^{(r,s)},\mJ^{(r,s)}) = (\mJ^{(0,0)},\mJ^{(0,0)})  + r^2 +  7 s^2 - 5r\, s  \,.
\end{eqnarray}
One possibility is $r=1,s=0$, which changes the self-intersection by $1$.
This which produces the $(1,0)$ adjoint with $(-1)$ self-intersection (except
for the Cartan generators) from the $(0,0)$ roots with non-Cartan
self-intersection $(-2)$.  A second transformation is $r=2, s=1$,
which also changes $(\mJ, \mJ)$ by 1, resulting in
another 240 junctions of self-intersection $(-1)$ at charge $(2,1)$.  We
could also have obtained these directly from $(1,0)$ with $r=1,s=1$,
which does not change the self-intersection, as required.  These have been
examples to illustrate the idea; many more sets of junctions with
self-intersection $(-1)$ should exist.

\section{Dynkin labels of string junctions}\label{s:Dynkin}

For an arbitrary string junction associated to a collection of branes with
some gauge symmetry ${\cal G}$, we would like to determine 
the corresponding weight vector
$\vec{\lambda}$ giving the charge assignments of the string state.  Hence we seek
to determine the Dynkin labels associated to the junction in terms of the invariant 
charges:
\begin{equation}
\label{afromQ}
a_i(\mJ) = Q^\mu(\mJ)\, K_{\mu i} \,,
\end{equation}
where $i = 1, \ldots , {\rm rank} \, {\cal G}$, and $\mu$ indexes the various
branes.   The purpose of the present section is to determine the matrix $K_{\mu i}$
and  we will follow an explicit  approach in doing so for the case of 
$su(n)$ and $so(8)$.  We will exhibit junctions for the fundamental representations
and discuss the interplay between conjugacy classes of representations and 
asymptotic charges.  The strings representing the 
${\bf 8_v},{\bf 8_s}$ and  ${\bf 8_c}$ representations
of $so(8)$ were derived independently in \cite{imamura}.

In the next section we will actually show that the results of the present section
can be reproduced from the  intersection bilinear. We
will see that  $a_i(\mJ) = - \left( \mJ  , \malpha_i \right) $.
Thus the Lie algebra properties of a string junction are encoded in its
intersections with the simple roots.

\subsection{Dynkin labels of su(n)}\label{ss:Dynkinsu}

We begin the analysis by focusing on the adjoint representation.
The root generators consist of strings that stretch between a pair of branes.
Since the Cartan matrix gives the Dynkin labels of the simple roots, 
we have from (\ref{afromQ}):
\begin{equation}
\label{yu1}
a_i(\malpha_j) = A_{ji} 
= Q^\mu (\malpha_j) K_{\mu i} \equiv Q_{j\mu} K_{\mu i} \,,
\end{equation}
which is a constraint on $K_{\mu i}$.  
Note that the index $\mu$ runs over one more value than the index $i$.
In order to solve the constraint effectively we do a change of basis.
We introduce new charges  $\{ \widehat Q^1 ,
\cdots  \widehat Q^{n-1}, Q^*\}$ with $\widehat Q^i = Q^i - Q^{i+1}$,
and 
\begin{eqnarray}
\label{pfromQsun}
Q^* = \sum_{i=1}^{n} Q^i = p \, , 
\end{eqnarray}
where $p$ denotes the total asymptotic charge, and is also 
the $u(1)$ charge in the $su(n)\times u(1) \subset u(n)$.
Since the roots carry $Q^*=0$, eq.~(\ref{yu1}) in the new basis reads
\begin{equation}
\label{yu2}
A_{ji} = \widehat Q^k (\malpha_j)\,  \widehat  K_{k i} \equiv 
\widehat Q_{jk} \widehat K_{k i} \,.
\end{equation}
The matrix  $\widehat Q_{jk} = \widehat
Q^k (\malpha_j)$, on account of $\malpha_i = \ma_i - \ma_{i+1}, \, 
i= 1, \ldots, n-1$,  is seen to be precisely equal to the Cartan matrix
$A_{jk}$. Thus 
$\widehat K_{ij}$ is
the identity matrix, and we therefore have
\begin{equation}
\label{hjgg}
a_i = \widehat Q^\mu \,\widehat K_{\mu i} = \widehat Q^i  +
Q^* \, k_i = Q^i - Q^{i+1}  + Q^* \, k_i\,, \,
\quad i=1, 
\ldots ,  n-1.
\end{equation}
The constants $k_i$ are determined by consideration of other
representations. A single string beginning on the configuration and 
going to infinity has $n$
choices of which brane to begin on, and transforms 
in the fundamental ${\bf n}$ of $su(n)$. These states have
$Q^*=1$.  Identifying the highest
weight vector $(1,0,\cdots , 0)$ of the fundamental
with an outgoing string from the 
first $A$ brane (no other possibility is consistent)
we find immediately that $k_i=0$. 
We therefore get 
\begin{equation}
\label{hjg}
a_i =  Q^i - Q^{i+1}  \,, \, \quad i=1,  \ldots ,  n-1.
\end{equation}
The  antifundamental is realized by strings ending on the branes
($Q^* = -1$).  Note that the fundamental and antifundamental representations
are composed precisely of the junctions with self-intersection $(-1)$.

\medskip
We can now consider the reverse problem, finding the charges defining the
junction given the Dynkin labels of a state.  
Since there is one more
charge than Dynkin label eq.~(\ref{hjg}) must be supplemented by 
equation (\ref{pfromQsun}) where we think of $p$ as an extra Dynkin
label. Solving for the $Q^i$'s in terms of the $\{a_i, p\}$ we find
\begin{equation}
Q^i = \sum_{r=i}^{n-1} a_r + \frac{1}{n} \Bigl( p - \sum_{r=1}^{n-1} \: r \: a_r\Bigr) . 
\end{equation}
We discover that for a given weight vector, $p$
must have a certain value mod $n$ in order to assure the $\{Q^i\}$ are
integral:
 \begin{eqnarray}
p &=& \sum_{r=1}^{n-1} r\ a_r\:\: ({\rm mod}\ n)  \,. \label{sunconj}
\end{eqnarray}
The right hand side of (\ref{sunconj}) is a constant 
for every state in a $su(n)$ representation. In fact, its value labels which
of the $n$ conjugacy classes of $su(n)$ the representation belongs to. 
Thus we find an important constraint:
the asymptotic charge $p= Q^*$ of a junction is fixed (mod $n$) by the 
conjugacy class of the representation it transforms under. 
As we shall see, the correspondence between $u(1)$ factors and
asymptotic charges, and the constraints relating them to conjugacy
classes,  is generic and  will persist in the more complicated algebras 
realized on mutually nonlocal branes.

\begin{figure}
$$\BoxedEPSF{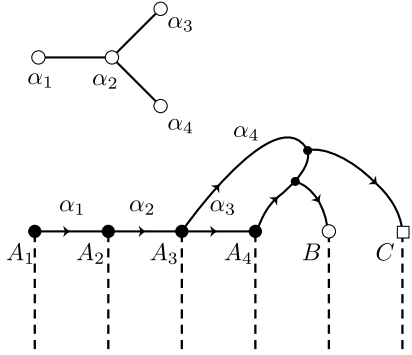 scaled 800}$$
\caption[Dynkin diagram and branes for $so(8)$]{Dynkin diagram for $so(8)$ and
associated brane configuration with junctions representing the simple roots.}
\label{dynkinso8}
\end{figure}

\subsection{Dynkin labels of so(8)}\label{ss:Dynkinso8}

The $so(8)$ algebra can be realized with four A branes, a B brane and a C
brane.  The Dynkin diagram and the brane configuration are shown in
Fig.~\ref{dynkinso8}.  The algebra
manifests itself through the regular maximal subalgebra
\begin{equation}
\label{decso8}
so(8) \ra su(4) \times u(1) \,,
\end{equation}
where $su(4)$ is the manifest semisimple symmetry of the branes and 
arises in the Dynkin diagram by removing the root $\vec\alpha_4$,
whose associated weight we conventionally chose to be the highest weight of
the ${\bf 8_c}$ representation. We note a few useful decompositions: 
\begin{eqnarray}
\label{decadj}
{\bf 28} &\ra  &{\bf 15}_0 \oplus {\bf 1}_0 
\oplus {\bf 6}_2 \oplus {\bf 6}_{-2}\,, \nonumber \\
{\bf 8_v} &\ra & {\bf 4}_1 \oplus \overline {\bf 4}_{-1} \,, \\
{\bf 8_s} &\ra & {\bf 4}_{-1} \oplus \overline {\bf 4}_{1}  \,,  \nonumber \\
{\bf 8_c} &\ra & {\bf  6}_{0 } \oplus  {\bf 1}_{2} \oplus  {\bf 1}_{-2} \,. \nonumber  
\end{eqnarray}
The string junctions making up the roots of $so(8)$
are the twelve roots of
$su(4)$ given in (\ref{sondirectroots}),  
and  strings which wind around the B and C branes in  
traveling between A-branes. These are the roots in (\ref{sonindirectroots}).

The $u(1)$ factor in (\ref{decso8})  is generated by
\begin{eqnarray}
\label{so8u1}
H^* &=& H_1 + H_2 + H_3 + H_4 \,,
\end{eqnarray}
where $H_i$ is the $i$th Cartan generator of $so(8)$.  One can check that
$H^*$ indeed commutes with all the generators of the $su(4)$:
\begin{eqnarray}
[ H^*, E_\alpha ] = \left( \sum_{i=1}^4 \alpha^i \right) E_\alpha =
\vec{\alpha} \cdot \left( \sum_{i=1}^4 \vec{e}_i \right) E_\alpha =
2 \left( \vec{\alpha} \cdot \vec{\omega}^4 \right) E_\alpha = 0 \, ,
\end{eqnarray}
where we have used
$2\vec{\omega}^4 = \vec{e}_1 + \vec{e}_2 + \vec{e}_3 + \vec{e}_4$,
and (\ref{fweights}). 

\medskip
We now relate Dynkin labels to the invariant
charges.  This time we note that there are two more branes than Dynkin
labels. At the same time we now have two nontrivial asymptotic
charges $p$ and $q$ given by 
\begin{eqnarray}
p &=& \sum_{i=1}^n Q_A^i + Q_B + Q_C \,.\label{sopq} \\
q &=& -Q_B + Q_C \,.  \nonumber
\end{eqnarray}
with $n=4$.
While for the $u(n)$ configurations
of the previous subsection $p$ denoted the $u(1)$ generated by all the $A$ 
branes,  here $p$ and $q$ are two linear combinations of three $u(1)'s$; the $u(1)$
generated by the all the A branes, the $u(1)$ of the B brane, and the 
$u(1)$ of the C brane.  

This time we have $a_i = Q^\mu \, K_{\mu i}$ with $\mu$ running over
six values while $i$ runs over four values.  Just as in the $su(n)$ case
we change basis from the six original charges to charges
$\{ \widehat Q^1_A ,\cdots \widehat Q_A^{4}, p, q\}$ defined by 
\begin{eqnarray}
\widehat Q^i_A &=& Q^i_A - Q_A^{i+1} \quad i = 1,2,3, \\
\widehat Q^4_A &=& Q^3_A + Q^4_A. \nonumber
\end{eqnarray}
For the roots $\malpha_j$, $p=q=0$, and thus
$a_i(\malpha_j) = \widehat Q^\mu(\malpha_j) \, \widehat K_{\mu i}$ becomes
$A_{ji} = \widehat Q^k (\malpha_j)\,  \widehat K_{k i} \equiv \widehat Q_{jk}
\, \widehat K_{k i}$ in this basis. 
Once more, explicit consideration of the roots shows  
that $Q_{jk} = A_{jk}$ and thus $\widehat K_{ki}$ is the identity matrix.  We
therefore have 
\begin{eqnarray}
a_1 &=& Q_A^1 - Q_A^2 \,, \nonumber \\
a_2 &=& Q_A^2 - Q_A^3 \,, \label{aso8first} \\
a_3 &=& Q_A^3 - Q_A^4\,,  \nonumber \\
a_4 &=& Q_A^3 + Q_A^4 + \theta p + \kappa q \nonumber\,,
\end{eqnarray}
where we have included the possibility of $p$ and $q$ contributions
for $a_4$ but not for the other labels.  In fact such contributions
necessarily vanish. The first three Dynkin labels are those of the $su(4)$
subalgebra and we already know that (\ref{aso8first}) gives the correct values
for that case. 
Moreover,  
B and C strings must be $su(4)$ singlets and $p,q$ contributions
to the first three Dynkin labels would ruin this. 

\smallskip
It only remains to determine the constants $\theta$ and $\kappa$.
We can do so by
considering the three inequivalent ${\bf 8}$ representations, whose highest
weight vectors are given by
\begin{eqnarray}
\label{wvfund}
{\bf 8_v} : (1,0,0,0) \,, \quad
{\bf 8_s} : (0,0,1,0) \,, \quad
{\bf 8_c} : (0,0,0,1) \,.
\end{eqnarray}
We have found in the previous section precisely three groups of
eight junctions with self-intersection $(-1)$. 
We will show that fitting them into the three available ${\bf 8}$
representations can
only be done in a unique way. Consider first the states with $(p,q) =
(1,0)$, which in view of (\ref{aso8first}) imply 
\begin{eqnarray}
\label{ffirst}
a_i &=& Q_A^i - Q_A^{i+1}\,, \quad i= 1,2,3,  \nonumber \\
a_4 &=& Q_A^3 + Q_A^4 + \theta \, .
\end{eqnarray}
The eight states were listed in (\ref{strs1}). Suppose we try to fit them to
${\bf 8_c}$. This is clearly impossible since we have a single
nonvanishing $Q_A$ and the first three zeroes in the the $(0,0,0,\pm 1)$
weights
cannot be obtained. Now let us try to fit them into the ${\bf 8_s}$, in
particular
focus on the $(0,0,\pm 1,0)$ weights. Again since only one $Q_A$ is
nonvanishing,
the first two zeroes imply that it must be $Q_A^4$. The junction $\ma_4$
could only
represent $(0,0,-1,0)$ requiring $\theta = -1$. But then, the only
junction that
could possibly represent $(0,0,1,0)$ is $-\ma_4 + \mb + \mc$ and it does not.
The only consistent possibility is to assign the eight junctions to the
${\bf 8_v}$,
in which case $\ma_1$ must give the $(1,0,0,0)$, fixing $\theta =0$.

\begin{figure}
$$\BoxedEPSF{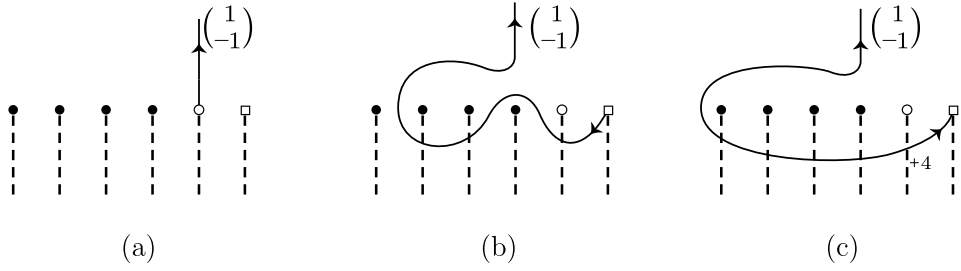 scaled 900}$$
\caption[Alternative junctions for ${\bf 8_c}$]{String junctions with asymptotic
charge $(1,-1)$. (a) The string $\mb$. (b)
The string $(-\ma_2-\ma_3+2 \mb + \mc )$. (c) The string
$(-\sum \ma_i + 3\mb + 2\mc)$. The strings in (a) and (c)
are singlets under $su(4)$.  The strings shown  belong to the ${\bf 8_c}$.}
\label{string8cb}
\end{figure}

Let us now try to fit another group of eight states to the ${\bf 8_c}$
representation.
We see that no states in (\ref{strs2}) can represent  
the weights $(0,0,0,\pm 1)$. The two states in (\ref{lone}), however, can. 
This time, with $(p,q)= (1,1)$ we have
$a_4 = Q_A^3 + Q_A^4 + \kappa$. If we tried to identify $\mc$ with $(0,0,0,1)$
we need $\kappa=1$, but then the junction 
$\sum \ma_i -2\mb -\mc$ cannot give
$(0,0,0,-1)$. The other way around it works, we identify $\mc$ with
$(0,0,0,-1)$
fixing $\kappa=-1$. Then the junction $\sum \ma_i -2\mb -\mc$  gives
$(0,0,0,-1)$.  The complete formulas for the Dynkin labels of $so(8)$ are
 therefore
\begin{eqnarray}
\label{aso8}
a_1 &=& Q_A^1 - Q_A^{2} \, ,\nonumber\\
a_2 &=& Q_A^2 - Q_A^{3}\,, \nonumber\\
a_3 &=& Q_A^3 - Q_A^{4}\,, \\ 
a_4 &=& Q_A^3 + Q_A^4 + Q_B - Q_C\, . \nonumber
\end{eqnarray}
The $u(1)$ charge, from Eqn.~(\ref{so8u1}), is
\begin{eqnarray}
Q^* = a_1 + 2a_2 + a_3 + 2a_4 = 
\sum_{i=1}^4 Q_A^i + 2\, Q_B - 2\, Q_C. \label{u1so8}
\end{eqnarray}
With these formulae, the ${\bf 28}$, ${\bf 8_v}$, ${\bf 8_s}$, and
${\bf 8_c}$ can all be represented as simple strings that trace Jordan
curves.  The asymptotic
charges associated to these strings are
\begin{eqnarray}
\label{gotrep}
{\bf 28}: \left( 0 \atop 0 \right) \,, \quad  
{\bf 8_v}: \left( 1 \atop 0 \right) \,,
\quad {\bf 8_s}: \left( 0 \atop 1 \right) \,, \quad
{\bf 8_c}: \left( 1 \atop 1 \right) \,.
\end{eqnarray}
Examples were shown in Figs.~\ref{string8v},~\ref{string8s},~\ref{string8c}.  
In Fig.~\ref{string8v}, the first string belongs to the ${\bf 4}_1$, and the
second belongs to the  ${\bf\bar{4}}_{-1}$.  In Fig.~\ref{string8s}
the first string belongs to the 
${\bf 4}_{-1}$ and the second to the ${\bf \bar{4}}_1$.  
In Fig.~\ref{string8c}  we have states in  the ${\bf 8_c}$ with
asymptotic charge $(1,1)$. The first shown, a C string,  is in the ${\bf
1}_{-2}$; the second one  is a representative from the ${\bf 6}_0$ and 
the third  is the ${\bf 1}_2$.  The reader may wonder about the symmetry
of strings departing B and strings departing C. For this purpose we
have included Fig.~\ref{string8cb} showing states in the ${\bf 8_c}$
representation generated by strings with asymptotic charge $(1,-1)$.
The first string is the B string, in the
${\bf 1}_2$, the second is a representative of ${\bf 6}_0$ and the third is the
${\bf 1}_{-2}$.  These junctions are obtained from the $(1,1)$ ${\bf 8_c}$
via a shift by $\mathbold{\eta_2}$.  

\medskip
Equation (\ref{gotrep}) is consistent with familiar 
results in four dimensional 
${\cal N}=2$ supersymmetric $su(2)$ gauge theory with
four flavors \cite{sw}.   As is
well-known \cite{bf, bds, senthreebrane,fayyazuddin}, this theory can be realized
on the worldvolume of a D3-brane probe moving in the 
AAAABC background, with matter states in the worldvolume theory 
arising from  strings which begin on 7-branes and end
on the D3-brane.  In agreement with (\ref{gotrep}), 
in addition to $\ch{1}{0}$ elementary hypermultiplets transforming in the ${\bf
8_v}$ of $so(8)$, monopole and dyon states with  charges $\ch{0}{1}$ and
$\ch{1}{1}$, transforming in the ${\bf 8_s}$ and
${\bf 8_c}$ respectively were found \cite{sw}.  
In general an arbitrary
$\ch{p}{q}$ dyon was said to transform in the eight-dimensional representation
associated to  $\ch{p'}{q'}$, with $p'=p$ (mod 2) and $q'=q$ (mod 2). 
We explain this now.
Since the map from invariant charges to labels $\{a_i ,
p, q\}$ is invertible, as we explore in the next subsection, there must be two transformations of the
invariant charges that change $p$ and $q$ while preserving the Dynkin labels. 
These are easily found by inspection. One such
transformation is
\begin{equation}
\label{trone}
\delta Q_B =\delta Q_C =m\, ,
\end{equation}
with $m$ an arbitrary integer. This changes the asymptotic charge by 
$\ch{2m}{0}$. The other operation is 
\begin{equation}
\label{trtwo}
\delta Q_A^i = - n\ \forall i,\quad \delta Q_B = -3 n,\quad  \delta Q_C = - n\,,
\end{equation}
with $n$ an arbitrary integer. This changes the asymptotic 
charge $\ch{0}{2n}$. The reader may recognize that these transformations
are nothing else than the transformation (\ref{chla}) generated by
$\mathbold{\eta}_1$ and $\mathbold{\eta}_2$ given in (\ref{nullj}). 
In the next section we will see that these junctions do not change Dynkin labels
since they have zero intersection number with the roots. Given a
$\ch{p}{q}$ dyon we can therefore
soak the asymptotic charge $\ch{2m}{2n}$
necessary to reduce it 
 to one of the four canonical asymptotic charges 
via a transformation of the form (\ref{nullj}).  

\subsection{Conjugacy classes for so(8)}\label{ss:Conjugacy}

The conjugacy classes of $so(8)$ are labeled by two integers (mod 2):
\begin{eqnarray}
\label{so8con}
C_1 &=& a_3 + a_4 \quad (\hbox{mod}\, 2) \,, \label{conjso8}\\
C_2 &=& a_1 + a_3  \quad (\hbox{mod}\, 2) \,. \nonumber
\end{eqnarray}  
It follows from (\ref{aso8}) that 
\begin{eqnarray}
Q_A^1  &&= \fracs12\, ( 2a_1 + 2a_2 + a_3 + a_4 - q) \nonumber  \\
Q_A^2  &&= \fracs{1}{2} \, ( 2a_2 + a_3 + a_4 - q) \nonumber  \\
Q_A^3  &&= \fracs{1}{2} \,  ( a_3 + a_4 - q)   \\
Q_A^4  &&= \fracs{1}{2} \,( -a_3 + a_4 - q) \nonumber  \\
Q_B &&= \fracs{1}{2} \, ( -a_1 - 2a_2 - a_3 - 2a_4 +p+ q)  \nonumber \\
Q_C &&= \fracs{1}{2}\, ( -a_1 - 2a_2 - a_3 - 2a_4 +p+ 3q). \nonumber  
\end{eqnarray} 
The condition that the invariant charges must be integers requires
$q= a_3 + a_4 $ (mod 2), and $p= a_1 + a_4$ (mod 2),  and using (\ref{so8con})
we have   
\begin{eqnarray}
p \, (\hbox{mod}\, 2) &=& C_1 + C_2 \,,   \label{pqconjso8}\\
q \, (\hbox{mod}\, 2) &=& C_1\, .   \nonumber
\end{eqnarray} 
This shows that  the
conjugacy class of an $so(8)$ representation determines the required
asymptotic $\ch{p}{q}$ charge mod 2. The representation can be constructed
by junctions of any fixed asymptotic charge satisfying the mod 2 condition.
This is analogous to the situation for $su(n)$, where the conjugacy class
determined the possible values of $p \ (\hbox{mod}\, n)$. 
The above equation implies that $(C_1,C_2) \leftrightarrow \ch{p}{q}$ 
are related as follows
\begin{eqnarray}
(0,0) \leftrightarrow \left( 0 \atop 0 \right) \,, \quad 
(0,1) \leftrightarrow \left( 1 \atop 0 \right) \,, \quad
(1,1) \leftrightarrow \left( 0 \atop 1 \right) \,, \quad
(1,0) \leftrightarrow \left( 1 \atop 1 \right) \,, \label{ccpqso8}
\end{eqnarray} 
which can be checked to be consistent with the results
of  (\ref{gotrep}) for the 
${\bf 28}$,
${\bf 8_v}$,
${\bf 8_s}$ and ${\bf 8_c}$ representations.

\smallskip
Note that conjugacy classes of representations
form a group under tensor product of representations. 
Indeed, in the present example, if we take any representation of the 
conjugacy class
$(C_1, C_2)$ and we tensor it with any representation in the conjugacy
class $(C_1', C_2')$ the result is a series of representations all of 
which belong to the conjugacy class $(C_1, C_2) \cdot (C_1', C_2') =
(C_1+ C_1' , C_2 + C_2')$ (mod 2). Thus conjugacy classes of representations
combine according to the group $\bbbz_2 \times \bbbz_2$.
The above result associates to each conjugacy class of representations
an equivalence class of possible $(p,q)$ asymptotic charges. 
Since multiplication of representations implies 
addition of asymptotic charges for the corresponding
junctions, consistency requires that the map from conjugacy classes to 
equivalence classes of $(p,q)$ charges define a homomorphism from
the group of conjugacy classes to a group generated by the equivalence
classes of asymptotic charges under addition. Note that for our 
particular example we actually have an isomorphism of groups.  

\subsection{Dynkin labels of so(2n)}\label{ss:Dynkinso2n}

The construction of $so(8)$ generalizes readily to the case of $D_n = so(2n)$,
with the slight complication that the conjugacy classes are different
for $n$ odd than for $n$ even. 

Since the analysis is so similar to that of $so(8)$, we directly give the
result expressing Dynkin labels in terms of invariant charges:

\begin{eqnarray}
a_i &=& Q^i_A - Q^{i+1}_A \,, \quad i=1 \ldots n-1  \,, \label{aso2n}\\
a_n &=& Q_A^{n-1} + Q_A^n + Q_B - Q_C. \nonumber
\end{eqnarray}
The $u(1)$ charge is given by
\begin{eqnarray}
Q^* = \sum_{i=1}^{n-2} 2i\ a_i + (n-2)\ a_{n-1} + n\ a_n.
\end{eqnarray}
When $n$ is even one may  insert a factor of $\half$ to the definition of $Q^*$
to normalize the smallest value of $Q^*$ to unity.
Solving for the invariant charges we find
\begin{eqnarray}
Q^i_A &=& \fracs12 
\Bigl(\,\sum_{j=i}^{n-2} 2a_j + a_{n-1}+ a_n  + q \Bigr) \label{qaso2n} \,, \\
Q_B &=& - 
\fracs14 \Bigl( \sum_{j=1}^{n-2} 2j\ a_j + (n-2)\ a_{n-1} + n\ a_n -2p  +(n+2)\ q \Bigr) \,, \label{qbso2n} \\  
Q_C &=& - 
\fracs14 \Bigl( \sum_{j=1}^{n-2} 2j\ a_j + (n-2)\ a_{n-1} +
n\ a_n -2p  +(n-2)\ q \Bigr). \label{qcso2n}
\end{eqnarray}
These equations will place constraints on the asymptotic charges differently
for $n$ even and $n$ odd.
This is consistent with the different conjugacy classes of $so(2n)$ for
$n$ even and $n$ odd.  For $n$ even,  the group of conjugacy classes 
is $\bbbz_2 \times \bbbz_2$, as given by
\begin{eqnarray}
C_1 &=& a_{n-1} + a_n \quad (\hbox{mod}\, 2) \label{conjso2n}\\
C_2 &=& a_1 + a_3 + \cdots + a_{n-1}  \quad (\hbox{mod}\, 2) \,, \nonumber 
\end{eqnarray}  
which generalizes (\ref{conjso8}).  One can readily check that
the constraints on the
asymptotic charges are still given by (\ref{pqconjso8}), (\ref{ccpqso8}).

For $n$ odd, the situation is more complicated.  Conjugacy is $\bbbz_4$, given
by 
\begin{eqnarray}
C_2 &=& 2a_1 + 2a_3 + \cdots + 2 a_{n-2} + a_{n-1} - a_n
\quad {\rm (mod\ 4)} \,.
\end{eqnarray} 
Unlike the case for $n$ even, the quantity
\begin{eqnarray}
C_1 &=& a_{n-1} + a_n \quad {\rm (mod\ 2)} \,,
\end{eqnarray}
is not independent, but is determined by the value of $C_2$.  However, we
still find it useful to define since it characterizes the
constraint on the asymptotic charges.  We see that (\ref{qaso2n}) requires
\begin{eqnarray}
q &=& C_1 \quad {\rm (mod\ 2)} \,.
\end{eqnarray}
However, in (\ref{qbso2n}) and (\ref{qcso2n}) it is the value of $q$ (mod 4) that is relevant.
Thus $q = C_1$ (mod 4) and $q = C_1 + 2$ (mod 4) will correspond to
different conjugacy classes.  The value of $p$ in turn changes to compensate
for the different value of $q$, but since only $2p$ appears, $p$ is only
determined (mod 2).  The relationship is
\begin{eqnarray}
2p -q &=& C_2 \quad {\rm (mod\ 4)} \,.
\end{eqnarray}
This corresponds to the following map between conjugacy (given by $C_2$) and asymptotic charges, (mod 2) for $p$ and (mod 4) for $q$:
\begin{eqnarray}
0 \leftrightarrow \left( 0 \atop 0 \right), \left( 1 \atop 2 \right) \,, \quad 
1 \leftrightarrow \left( 1 \atop 1 \right), \left( 0 \atop 3 \right) \,, \quad
2 \leftrightarrow \left( 1 \atop 0 \right), \left( 0 \atop 2 \right) \,, \quad
3 \leftrightarrow \left( 0 \atop 1 \right), \left( 1 \atop 3 \right) \,. 
\end{eqnarray} 

\medskip

\section{Weight lattice from junction lattice}\label{s:Weight} 

In previous section, we considered the $A_n$ and 
$D_n$ algebras and derived the formula giving the
Dynkin labels associated to arbitrary junctions by explicit examination of
certain representations.  The formula (\ref{afromQ}) 
requires finding the matrix $K$. We showed that $K$ is fixed by requiring the
the expected Dynkin labels for the root junctions and for the junctions of 
self-intersection is$(-1)$.  In this section we will find a simple explanation
for this fact using the intersection matrix. (\ref{afromQ}) will be
reinterpreted 
$a_i (\mJ) = - ( \mJ , \malpha_i )$.  We will show that proving this
requires, in general, examination of at least two representations that
have nontrivial asymptotic charges. This better
understanding of the relation between junction 
and weight lattices will allow us to 
complete the discussion of the exceptional algebras.  Additionally,
we solve for the invariant charges in terms of
$\{a_i, p, q \}$, and show that in each case, the conjugacy class of the algebra places
restrictions on the asymptotic charges.

\subsection{The general argument}

To begin with, let us find the constraints on the matrix  $K$ arising
from the root junctions.  
Observe that the   Cartan matrix,  giving the Dynkin labels of the simple roots,
is realized geometrically as the intersection of simple roots:
\begin{eqnarray}
a_i(\malpha_j) = A_{ji} = - 
\left(\malpha_j, \malpha_i \right) = Q^\mu (\malpha_j)
\, K_{\mu i} \,, 
\end{eqnarray}
where we also made use of (\ref{afromQ})
for the case of the adjoint. 
If we expand $\malpha_j$  in the basis of strings we find 
\begin{eqnarray}
 - \left(\malpha_j, \malpha_i \right)
 = - \, Q^\mu (\malpha_j) \,
\left(\ms_\mu \,, \malpha_i  \right) \,,
\end{eqnarray}
and from the last two equations we deduce that
\begin{equation}
\label{findk}
0\, = \, Q^\mu(\malpha_j) \bigl( K_{\mu i} + \left(\ms_\mu\,,  \malpha_i
\right)\bigr) \,.
\end{equation} 
To find the general solution of the above equation consider
the set of vectors associated to all simple roots: 
$Q^\mu (\malpha_1) , \cdots , Q^\mu
(\malpha_r)$. Let $r_\mu^1, \cdots , r_\mu^n$ be a basis for
a set of vectors
orthogonal to all of the $Q$ vectors, namely  
\begin{equation}
\label{exiu}
Q^\mu (\malpha_j ) r_\mu^l = 0\,,
\end{equation}
for all $j$ and $l$.  Then the general solution of (\ref{findk}) is
\begin{equation}
\label{gensol}
 K'_{\mu i}  =  -\left(\ms_\mu\,,  \malpha_i
\right) + \sum_{l=1}^n  \beta_{i\, l} r^l_\mu \, .
\end{equation}
The simplest solution is therefore
\begin{equation}
\label{simsol}
 K_{\mu i}  =  -\left(\ms_\mu\,,  \malpha_i
\right) \, . 
\end{equation}
If we take this to be the correct value for the matrix $K$ we can now
find a simple expression  for the Dynkin labels of junctions. We have
\begin{eqnarray}
\label{solsol}
a_i(\mJ) &=& Q^\mu (\mJ) K_{\mu i } = - Q^\mu (\mJ)\left( \ms_\mu \,, \malpha_i, 
\right)  \\
         &=& - \left(\mJ \,, \malpha_i \right) \,, \nonumber
\end{eqnarray}
showing that  the Dynkin labels of the weight associated to any junction are
determined by its intersection with junctions associated to the simple roots.
This is exactly the result we said we would prove in (\ref{hui}). It was
said then that the constants $a_i$ in (\ref{jexp}) are indeed the Dynkin labels
of the associated weight vector.

We now explain why (\ref{solsol}) is the correct solution.
Assume we have a set of junctions $\{ \mJ_\alpha \}$  all of which
have the same asymptotic charges $(p,q)$. Similarly, assume we have another set 
of junctions $\{ \mJ'_\beta\}$ all of which have asymptotic charge $(p',q')$.
Assume further that both $(p,q)$ and $(p',q')$ are not identically zero, nor
proportional to each other. Finally, assume (\ref{solsol}) identifies correctly
these two sets of junctions to two representations. Under these circumstances,
we will show that (\ref{simsol}) is the only possible solution.

Assume that $a_i (\mJ_\alpha) = Q^\mu(\mJ_\alpha) K_{\mu i}$  maps
correctly the junctions $\{ \mJ_\alpha\}$ to
the full set of weight vectors
$\vec\lambda (\mJ_\alpha)$ of a specific representation $R_\alpha$.  
Note  that from (\ref{gensol}) it
follows that 
\begin{equation}
\label{getrid}
a_i'(\mJ_\alpha) = a_i (\mJ_\alpha ) + \sum_{\beta, \mu} \beta_{il}\, 
r^l_\mu Q^\mu (\mJ_\alpha) \, .
\end{equation}
We now claim that there are two vectors $r_\mu$ satisfying the required
orthogonality condition (\ref{exiu}). These are the vectors 
$r_\mu^1$ and $r_\mu^2$ satisfying for arbitrary junctions
\begin{equation}
\label{needpq}
Q^\mu (\mJ) r_\mu^1 = p(\mJ) , \quad Q^\mu (\mJ) r_\mu^2 = q(\mJ)\,.
\end{equation}
Since root junctions have zero asymptotic charges the requisite conditions
 (\ref{exiu}) are satisfied. Now eq.~(\ref{getrid}) can be rewritten as
\begin{equation}
\label{getridd}
a_i'(\mJ_\alpha) = a_i (\mJ_\alpha ) + \beta_{i} \, p (\mJ_\alpha)
+  \gamma_{i} \, q (\mJ_\alpha) \, .
\end{equation}
The important thing to note here is that since all junctions
have the same asymptotic charges all weight vectors $\vec\lambda
(\mJ_\alpha)$ are shifted by the same vector $\vec\rho_0$; namely 
\begin{equation}
\label{bnbn}
\vec\lambda' (\mJ_\alpha) = \vec\lambda (\mJ_\alpha) + \vec\rho_0\, ,
\end{equation}
where
\begin{equation}
\vec\rho_0 = \sum_i (\beta_i p + \gamma_i q) {\vec w}^i\, . 
\end{equation}
Since the set
of weights  of any (irreducible) representation $R_\alpha$ cannot be invariant under
translations we conclude that $\vec \rho_0 = \vec 0$. 
Because the fundamental weights
${\vec w}^i$ form a basis, if $\vec \rho_0 = \vec 0$ then 
\begin{equation} 
(\beta_i p + \gamma_i q)= 0\,, \forall  \, i . 
\end{equation}
Similarly, the second set of junctions would require 
$(\beta_i p' + \gamma_i q')= 0\,, \forall  \, i .$ Since we assumed that the
$(p,q)$ and $(p',q')$ values are nonzero and not proportional, this implies
that the constants $\beta_i$ and $\gamma_i$ vanish identically.
This concludes our proof. 

The above discussion is in agreement with our experience in
the previous section. For the $D_n$ algebras we had to examine
two sets of junctions with non-vanishing asymptotic charges
to fix the matrix $K$ (for the case of $A_n$ algebras, as usual,
one representation sufficed).

\subsection{A${}_{\bf n}$ and D${}_{\bf n}$ algebras revisited}

We now verify that our previous results for 
$su(n)$ and $so(8)$ follow directly from (\ref{solsol}).  
Consider first $su(n)$, and examine $a_i = - Q^\mu (\ms_\mu, \malpha_i)$.
Since $\malpha_i = \ma_i - \ma_{i+1}$, the formula gets contributions
only from $\mu= i, i+1$, reproducing $a_i = Q^i - Q^{i+1}$.

In a similar fashion we can confirm our results for $so(8)$.  The simple
roots are given in (\ref{sonsimple})
\begin{eqnarray}
\malpha_i = \ma_i - \ma_{i+1} \quad i=1,2,3 \,, \qquad
\malpha_4 =
\ma_3 + \ma_4 - \mb - \mc\,. 
\end{eqnarray}
For $i=1,2,3$ we find $a_i = Q^i_A - Q^{i+1}_A$ since the $\mb$ and $\mc$ 
strings have zero intersection with the first three root junctions. 
We then get $a_4 = - Q^\mu (\ms_\mu, \malpha_4)$ and the only
nontrivial terms are 
\begin{eqnarray}
a_4 &=&  - Q^3_A\, (\ma_3\, , \malpha_4)- Q^4_A\, (\ma_4\, , \malpha_4)
- Q_B\,  (\mb\, , \malpha_4)- Q_C\, (\mc\, , \malpha_4)\,, \nonumber\\
&=&   Q^3_A+ Q^4_A  + Q_B - Q_C\,,
\end{eqnarray}
reproducing 
(\ref{aso8}).  The case of $so(2n)$ proceeds along the same lines.

\subsection{Matching junctions to special representations}

Looking for junctions of self-intersection $(-1)$ for the exceptional algebras,
we found sets of 27, 56 and 240 junctions for $E_6$, $E_7$ and $E_8$
respectively.  How can we confirm that (\ref{solsol})
associates these junctions precisely 
to the weight vectors of the ${\bf 27}$, the ${\bf 56}$, 
and to the non-zero weights of the ${\bf 248}$? We claim that
it is sufficient to prove that for each case one junction maps
to a weight vector of the representation in question.
This will be simple to show, given that all weights of ${\bf 27}$,
all weights of the ${\bf 56}$, 
and all non-zero weights of the ${\bf 248}$ have the 
same length, and no weight appears more than once in the
representations.\footnote{The same is true for the vector and spinor
representations of the $D_n$ series, as well as for the
${\bf \overline{27}}$ of $E_6$. We thank V. Kac for bringing these facts
to our attention.}

Focus on one of the algebras and the set of $(-1)$ self-intersection
junctions that we wish to show is associated to a representation $R$.
Assume there is a junction $\mJ$ in the set that maps to a weight vector
$\vec\lambda$ in $R$. Let $\vec\alpha$ be a simple root that can be added to 
$\vec\lambda$ to get another weight vector $\vec \lambda' = \vec\lambda +
\vec\alpha$ in $R$. The linearity of (\ref{solsol}) implies that $\mJ' = \mJ + 
\malpha$ is mapped to $\vec\lambda'$. We now want to show that $\mJ'$ and
$\mJ$ have the same self-intersection. Recall that assuming (\ref{solsol})
we showed that the self-intersection of a junction $\mJ$ is related to the
length squared of its associated weight vector by
\begin{eqnarray}
\label{selfintlength}
- (\mJ, \mJ) = \vec{\lambda}(\mJ) \cdot \vec{\lambda}(\mJ) + f (p,q) \,,
\end{eqnarray}
where $f(p,q)$ is a quadratic form in the asymptotic charges, as 
(\ref{relcone6}), (\ref{relcone7}), (\ref{relcone8}). 
Since $\malpha$ has zero asymptotic charge, $\mJ$ and $\mJ'$ have the same
asymptotic charges and
\begin{equation}
\label{jfromlambda}
(\mJ', \mJ') - (\mJ, \mJ) = -\, (\vec\lambda' \cdot \vec\lambda' - 
\vec{\lambda} \cdot \vec{\lambda} ) \,. 
\end{equation}
Thus as long as $\vec\lambda'$ has the same length as $\vec\lambda$, $\mJ'$ 
has the same self-intersection as $\mJ$.  

The claim now follows easily. Starting with any weight produced
by a junction, we can by the process of adding or subtracting simple
roots reach every weight in the representation $R$. In the case of
$E_6$ and $E_7$,
for each such weight we find a junction of self-intersection $(-1)$ mapping to it. 
Since no weight appears more than once we produce a set of dim$(R)$ different
junctions of self-intersection $(-1)$ that give correct Dynkin labels.
Since, by assumption, there are precisely dim$(R)$ junctions of
self-intersection $(-1)$ for the given values of $(p,q)$ the set of junctions
produced by the iterative procedure must coincide with the junctions
we found in the lattice.  For the case of $E_8$ we note that the 240 nonzero
weights break into two groups, the positive roots and the negative roots.
All positive roots can be obtained from a positive root by adding and
subtracting simple roots, and the same holds for the negative roots.
If we have a junction mapping to a positive root, minus that junction
will map to the corresponding negative root. The above argument then 
can be applied to the two sets separately.  This concludes our proof. 

We now turn to a discussion of each exceptional algebra individually.

\subsection{Dynkin labels of E${}_{\bf 6}$}\label{ss:Dynkine6}

The simple roots of $E_6$ are given in (\ref{simpe6}).  
Compared to $so(8)$, there is now an additional A brane and an additional C
brane.

The roots like $\malpha_j = \ma_i - \ma_{i+1}$ produce cancellations
and give $a_j = Q^i_A - Q^{i+1}_A$ as before.  Similarly, we have
$\malpha_5 = \mc_1 - \mc_2$, and obtain $a_5 = Q_C^1 - Q_C^2$.  Finally,
$\malpha_4$ is analogous to $\malpha_4$ of $so(8)$.  We find
\begin{eqnarray}
a_1 &=& Q_A^1 - Q_A^2 \nonumber \\  
a_2 &=& Q_A^2 - Q_A^3 \nonumber \\
a_3 &=& Q_A^3 - Q_A^4 \label{e6dynkin} \\
a_4 &=& Q_A^4 + Q_A^5  + Q_B - Q_C^1 \nonumber \\
a_5 &=& Q_C^1 - Q_C^2 \nonumber \\
a_6 &=& Q_A^4 - Q_A^5. \nonumber
\end{eqnarray}
We invert the matrix $K$ to find for the invariant charges 
\begin{eqnarray}
Q_A^1 &=& 
\fracs{1}{3} \left( 4a_1 + 5a_2 + 6a_3 + 4a_4 + 2a_5 + 3a_6 -p + 3q
\right) \nonumber \,, \\ 
Q_A^2 &=& 
\fracs{1}{3} \left( a_1 + 5a_2 + 6a_3 + 4a_4 + 2a_5
+ 3a_6 -p + 3q \right) \nonumber \,, \\ 
Q_A^3 &=& \fracs{1}{3} \left( a_1 + 2a_2 + 6a_3
+ 4a_4 + 2a_5 + 3a_6 -p + 3q \right) \nonumber \,, \\ 
Q_A^4 &=& \fracs{1}{3} \left( a_1
+ 2a_2 + 3a_3 + 4a_4 + 2a_5 + 3a_6 -p + 3q \right) \,, \label{e6Q} \\
Q_A^5 &=& \fracs{1}{3} \left(
a_1 + 2a_2 + 3a_3 + 4a_4 + 2a_5  -p + 3q \right) \nonumber \,, \\ 
Q_B &=& \fracs{1}{3}
\left( -4a_1 -8a_2 -12a_3 -10a_4 -5a_5 -6a_6 +4p -9q \right) \nonumber \,, \\ 
Q_C^1
&=& \fracs{1}{3} \left( -2a_1 -4a_2 -6a_3 -5a_4 -a_5 -3a_6 +2p -3q \right)
\nonumber \,, \\ 
Q_C^2 &=& \fracs{1}{3} \left( -2a_1 -4a_2 -6a_3 -5a_4 -4a_5 -3a_6
+2p -3q \right) \,. \nonumber
\end{eqnarray}
Again the condition that $\{Q^i\} \in \bbbz$ requires that the charges
of a junction be determined by the conjugacy class of the representation.
$E_6$ has $\bbbz_3$ conjugacy determined by the quantity
\begin{equation}
C = a_1 - a_2 +a_4 - a_5 \quad {\rm (mod\ 3).}
\end{equation}
and we see that we must have
\begin{equation}
p = C \quad {\rm (mod\ 3)} \,. 
\end{equation}

The highest weight vector of the ${\bf 27}$ has Dynkin labels $(1,0,0,0,0,0)$.
Using (\ref{e6dynkin}), one can check that  the  junction $\ma_1$,  
present in the set of 27
junctions of  self-intersection $(-1)$, maps to this weight vector.  
Hence this set of 27 junctions with asymptotic charge
$(1,0)$  is associated to a ${\bf 27}$.  Additionally, we can use the transformation
(\ref{e6shift}) to produce another ${\bf 27}$, with asymptotic charges $(1,1)$.  Since
(\ref{solsol}) successfully identifies these two sets of junctions with $(p,q)$ not
proportional, by our previous arguments it is the only solution for $E_6$.

\subsection{Dynkin labels of E${}_{\bf 7}$}\label{ss:Dynkine7}

The simple roots of $E_7$ are given in Eqn.~(\ref{simpe7}).  There
is one more A-brane than in $E_6$.  The roots have the same basic
structure, and similar calculations lead us to the result
\begin{eqnarray}
a_1 &=& Q_C^1 - Q_C^2 \nonumber \,, \\  
a_2 &=& -Q_A^1 - Q_A^2  - Q_B + Q_C^2 \nonumber \,, \\
a_3 &=& Q_A^2 - Q_A^3 \nonumber \,, \\
a_4 &=& Q_A^3 - Q_A^4  \,, \label{e7dynkin}\\
a_5 &=& Q_A^4 - Q_A^5 \nonumber \,, \\
a_6 &=& Q_A^5 - Q_A^6 \nonumber \,, \\
a_7 &=& Q_A^1 - Q_A^2. \nonumber
\end{eqnarray}
As we did for $E_6$, we can invert $K$ to find
\begin{eqnarray}
Q_A^1 &=& 
\fracs{1}{2} \left( -2a_1 -4a_2 -4a_3 -3a_4 -2a_5 -a_6 -a_7  -p + 3q \right)
\nonumber \,, \\ 
Q_A^2 &=& \fracs{1}{2} \left( -2a_1 -4a_2 -4a_3 -3a_4 -2a_5 -a_6
-3a_7  -p + 3q \right) \nonumber \,, \\ 
Q_A^3 &=& \fracs{1}{2} \left( -2a_1 -4a_2
-6a_3 -3a_4 -2a_5 -a_6 -3a_7  -p + 3q \right) \nonumber \,, \\ 
Q_A^4 &=& \fracs{1}{2}
\left( -2a_1 -4a_2 -6a_3 -5a_4 -2a_5 -a_6 -3a_7  -p + 3q \right) \nonumber \,, \\
Q_A^5 &=& \fracs{1}{2} \left( -2a_1 -4a_2 -6a_3 -5a_4 -4a_5 -a_6 -3a_7  -p + 3q
\right) \,, \\ 
Q_A^6 &=& \fracs{1}{2} \left( -2a_1 -4a_2 -6a_3 -5a_4 -4a_5 -3a_6
-3a_7  -p + 3q \right) \nonumber \,, \\ 
Q_B &=&  3a_1 +6a_2 +8a_3 +6a_4 +4a_5
+2a_6 +4a_7  +2p +5q  \nonumber \,, \\ 
Q_C^1 &=&  2a_1 +3a_2 +4a_3
+3a_4 +2a_5+ a_6 +2a_7  +p -2q  \nonumber \,, \\ 
Q_C^2 &=& a_1 +3a_2
+4a_3 +3a_4 +2a_5 +a_6 +2a_7  +p -2q  \,. \nonumber
\end{eqnarray}
$E_7$ has ${\bbbz}_2$ conjugacy, given by
\begin{equation}
C = a_3 + a_4 + a_6 + a_7 \quad {\rm (mod\ 2)} 
\end{equation}
and indeed we see for the invariant charges to be integral we require
\begin{equation}
p + q = C \quad {\rm (mod\ 2).}
\end{equation}

The highest weight of the ${\bf 56}$ has 
 Dynkin labels $(0,0,0,0,0,1,0)$. 
As one can verify using (\ref{e7dynkin}),  the 
junction 
\begin{eqnarray}
-\sum_{i=1}^6 \ma_i -\ma_6 +4 \mb + 2 \sum_{j=1}^2 \mc_j \,,
\end{eqnarray}
with asymptotic charges $(1,0)$, maps to this weight vector.
This junction is among the set of 56
with self-intersection
$(-1)$, and implies that this set is identified with the   ${\bf 56}$.
(\ref{e7shift}) can be used to produce another ${\bf 56}$ with asymptotic
charges $(2,1)$, and these two representations together establish that (\ref{solsol}) is the
only solution for $E_7$ as well.

\subsection{Dynkin labels of E${}_{\bf 8}$}\label{ss:Dynkine8}
The simple roots of $E_8$ are given in Eqn.~(\ref{simpe8}).  There is now
a total of 7 A-branes.  Almost identical calculations give
\begin{eqnarray}
a_1 &=& Q_C^1 - Q_C^2 \nonumber \,, \\  
a_2 &=& -Q_A^1 -Q_A^2 - Q_B + Q_C^2 \nonumber \,, \\
a_3 &=& Q_A^2 - Q_A^3 \nonumber \,, \\
a_4 &=& Q_A^3 - Q_A^4 \,, \label{e8dynkin}\\
a_5 &=& Q_A^4 - Q_A^5 \nonumber \,, \\
a_6 &=& Q_A^5 - Q_A^6 \nonumber \,, \\
a_7 &=& Q_A^6 - Q_A^7 \nonumber \,, \\
a_8 &=& Q_A^1 - Q_A^2. \nonumber
\end{eqnarray}
Again we can invert $K$, to obtain for the invariant charges
\begin{eqnarray}
Q_A^1 &=& -2a_1 -4a_2 -5a_3 -4a_4 -3a_5 -2a_6 -a_7 -2a_8  -p +3q  \nonumber \,, \\ 
Q_A^2 &=& -2a_1 -4a_2 -5a_3 -4a_4 -3a_5 -2a_6 -a_7 -3a_8  -p +3q  \nonumber \,, \\
Q_A^3 &=& -2a_1 -4a_2 -6a_3 -4a_4 -3a_5 -2a_6 -a_7 -3a_8  -p +3q  \nonumber \,, \\
Q_A^4 &=& -2a_1 -4a_2 -6a_3 -5a_4 -3a_5 -2a_6 -a_7 -3a_8  -p +3q  \nonumber \,, \\
Q_A^5 &=& -2a_1 -4a_2 -6a_3 -5a_4 -4a_5 -2a_6 -a_7 -3a_8  -p +3q  \,, \\
Q_A^6 &=& -2a_1 -4a_2 -6a_3 -5a_4 -4a_5 -3a_6 -a_7 -3a_8  -p +3q  \nonumber \,, \\
Q_A^7 &=& -2a_1 -4a_2 -6a_3 -5a_4 -4a_5 -3a_6 -2a_7 -3a_8  -p +3q  \nonumber \,, \\
Q_B &=& 7a_1 +14a_2 +20a_3 +16a_4 +12a_5 +8a_6 +4a_7 +10a_8  +4p -11q  \nonumber \,, \\
Q_C^1 &=& 4a_1 +7a_2 +10a_3 +8a_4 +6a_5 +4a_6 +2a_7 +5a_8  +2p -5q  \nonumber \,, \\
Q_C^2 &=& 3a_1 +7a_2 +10a_3 +8a_4 +6a_5 +4a_6 +2a_7 +5a_8  +2p -5q . \nonumber 
\end{eqnarray}
$E_8$ has just one conjugacy class, and accordingly,  there is no 
condition on the asymptotic charges.

As discussed before, there are 240 junctions with asymptotic charge
$(1,0)$ and self-intersection $(-1)$.  One of these corresponds to the
highest weight $(0,0,0,0,0,0,1,0)$,
\begin{eqnarray}
-\sum_{i=1}^7 \ma_i -\ma_7 +8 \mb + 4 \sum_{j=1}^2 \mc_j \,,
\end{eqnarray}
and following our earlier arguments this is sufficient to establish that
these junctions are associated to the weights of the ${\bf 248}$ with
non-zero length.

Since the roots of $E_8$ have length squared 2 and the Cartan generators
have zero length squared, by (\ref{jfromlambda}) we expect the remaining 8 weights to
be associated to a junction with self-intersection $(+1)$.
The common weight vector has vanishing Dynkin labels, which requires $Q_A^i
\equiv Q_A \ \forall i$, $Q_C^1 = Q_C^2 \equiv Q_C$, and
\begin{eqnarray}
- 2 Q_A - Q_B + Q_C = 0 \,.
\end{eqnarray}
Furthermore we have for general asymptotic charges
\begin{eqnarray}
p &=& 7 Q_A + Q_B + 2 Q_C \,, \\
q &=& - Q_B + 2 Q_C \,. \nonumber
\end{eqnarray}
There is a unique solution for each $(p,q)$.  For our case of $(1,0)$ 
we find
\begin{eqnarray}
Q_A = -1 \,, \quad Q_B = 4 \,, \quad Q_C = 2 \,.
\end{eqnarray}  
One can readily check that the self-intersection of this
junction is indeed $(+1)$.

Using (\ref{e8shift}) we can produce another ${\bf 248}$, this one with
asymptotic charges $(2,1)$, and can use this ${\bf 248}$ together with
the previous one  to establish that (\ref{solsol}) is the only possible relation for $E_8$.
Thus we have proved that (\ref{solsol}) is
the unique solution for all simply laced Lie algebras.

\section{Conclusions and open questions}\label{s:Conclusion}

In this paper we have examined the Lie algebra representations arising
from string junctions ending on a background of branes.
On the space of junctions we have defined equivalence classes. Two
junctions belong to the same class if they can be transformed into
each other by brane crossings and continuous deformations. Each
equivalence class has a canonical representative, where the junction
does not cross the branch cut of any brane. 
The space of equivalence classes has the structure of a lattice, with
generators called basis strings. A general equivalence class is constructed
by integer linear combinations of basis strings.
This lattice is equipped with a metric, {\it i.e.} a bilinear symmetric inner product,
arising from intersection invariants.  This metric is positive definite for the
$A_n$ series, degenerate for the $D_n$ series, and nondegenerate but
indefinite for the $E_n$ series.

Additionally, we have  elucidated the relation between the lattice of junctions
on this background, and the weight lattice of the Lie algebra. 
For the $A_n$ series the junction lattice has one more dimension
than the weight lattice, while for the $D_n$ and $E_n$ series the junction
lattice has two more dimensions than the corresponding weight lattice.
In these latter cases, a choice of
asymptotic $(p,q)$ charges selects a junction sub-lattice of
codimension two which can be identified with the weight lattice of one 
conjugacy class of  representations of the algebra. A given weight vector in this
conjugacy class is thus represented as a junction with the chosen $(p,q)$. 
For any specific conjugacy class, the weight sub-lattice can actually
be identified with a set of $(p,q)$ junction sub-lattices. The set of $(p,q)$
values associated to each conjugacy class defines an equivalence class;  
this association being an isomorphism
between the group of conjugacy classes and the group of $(p,q)$ 
equivalence classes under addition.  The conjugacy class of the
adjoint maps to the equivalence class containing the asymptotic charge $(0,0)$.

Not only are junction and weight lattices related, but their metrics 
are related as well.
The identification of the $(0,0)$ junction sub-lattice to the root lattice
maps one metric into the other. For other
$(p,q)$ sublattices the two metrics differ by a quadratic form in $p$ and $q$.
Moreover, just as the Dynkin labels of a given weight are found by inner
product with the simple roots,  they are also given by the inner product
of the associated junction with the $(0,0)$ junctions representing the simple
roots.
It is useful to think of the entire junction space  as isomorphic to an
extended weight lattice, with additional fundamental weights  associated
to the asymptotic charges.
The additional weights are also related to the $u(1)$ generators of
the configuration that are not part of the enhanced gauge algebra.

Important questions remain. 
A full systematic analysis of the junction lattice, giving for example the 
states and representations for every possible intersection number, is missing. 
While in the standard weight lattice a 
given weight vector can appear in an arbitrary number of representations, 
one feels that the junction lattice, having more room in it, 
could organize states and representations in a less degenerate way.
Note that the junction lattice is in some sense more basic than the
lattice of homology classes of surfaces, the familiar object typically
used to see the emergence of Lie algebras (see \cite{aspinwall}). A basis
string, when lifted to the elliptic fiber,  is essentially half of a
two-sphere. We have not explored $F$-theory or $M$-theory viewpoints, nor
attempted to relate our work to the geometrical engineering program
\cite{leungvafa}.  The significance of the extended root system is not completely clear. 
As presently formulated, the extended Cartan matrix is block diagonal with
zero inner product between the new and old simple roots.  There may be other
relevant constructions. 

Our work has been largely Lie-algebraic,  and has not concentrated on the
string-theoretic realization of a junction.  In particular,
we have not attempted to determine which
representations of a given symmetry algebra have associated BPS junctions.
While non-BPS stable junctions are also of interest \cite{sennonbps},
BPS states are easily identified in the string spectrum and a full
understanding of their properties should be feasible.
It is known that two different holomorphic submanifolds 
must have a non-negative intersection number (see, for example
\cite{griffiths});  this condition, used in 
\cite{nekrasov}, could be of help in the present context as well. 
If a given representation is known to be realizable by BPS  junctions, it
remains to be shown that each state is uniquely realized by one and only one
representative in the corresponding equivalence  class of junctions for given
values of the moduli. These ideas could be relevant to elucidate
the physics of ${\cal N}=2$ four dimensional theories with exceptional flavor
symmetries.

By giving the general relation between junctions and weight vectors
we have made precise and extended in a significant way 
the usual picture of open strings stretched between branes as 
representing particular generators of a gauge algebra.  
Perhaps our methods will find 
further applications in other settings,
and help find generalized brane
constructions giving more exotic gauge algebras and representations.
\vskip.4in

\subsection*{Acknowledgments}

We would like to thank to T. Hauer and A. Iqbal for  many illuminating
discussions.  We are also grateful to M. Gaberdiel for a critical
reading of the manuscript and many useful suggestions, and   
to C. Vafa for helpful comments. Finally, we wish to thank M. Stock
for preparing the figures and for technical help with LaTeX.

\vskip.6in

\section*{Appendix: Dynkin labels for manifest subalgebras} 

In the brane configuration of $D_4$ the manifest subalgebra
(in the brane sense) is $su(4)\times u(1)$,  and for the brane
configuration of $E_{n+1}$, the
manifest subalgebra is $su(n)\times u(1)\times su(2)$.
Our goal here is to give the Dynkin labels of the manifest
subalgebras in terms of the Dynkin labels $\{ a_i\}$ of the enhanced algebras.
This is a standard embedding problem, specified by indicating
the relation of the simple roots
of the subalgebras  to the simple roots of full algebras. 

In all  of our cases  the manifest subalgebra has the same rank as the
enhanced algebra ${\cal G}$ and can be obtained from the
enhanced algebra through a series of maximal regular subgroups
\cite{johansen,GZ}.
Because the rank never changes, the sought-after transformation 
of Dynkin labels is nothing but a change of basis in weight space. 
The normalization of the $u(1)$ charge is conventional.

\smallskip
For $so(8) \ra su(4) \times u(1)$, as outlined previously, the
decomposition occurs by removing the fourth simple root and replacing it with the
$u(1)$ given in Eqn.~(\ref{u1so8}).  Letting
$\{b_i, Q^*\}$ denote the Dynkin labels of $su(4)$ and the $u(1)$ charge, we
find
\begin{eqnarray}
b_i &=& a_i\,, \quad i = 1,2,3, \nonumber \\
Q^* &=& a_1 + 2a_2 + a_3 + 2a_4 = \sum_{i=1}^4 Q_A^i + 2Q_B - 2Q_C. 
\end{eqnarray}

\smallskip
Consider now $E_6 \ra su(5) \times u(1) \times su(2)$. Letting
$\{b_i, Q^*, c\}$ denote the Dynkin labels of $su(5)$,  the $u(1)$
charge and the Dynkin label of $su(2)$ respectively, we find
\begin{equation}
(b_1,b_2,b_3,b_4)  = (a_1,  a_2,  a_3,  a_6), \quad  c = a_5 \,,
\end{equation}
\begin{eqnarray}
 Q^* &=& -4a_1 -8a_2 -12a_3 -10a_4 -5a_5 -6a_6 \nonumber\\
& = & -4\sum_{i=1}^5
Q_A^i -10Q_B +5\sum_{i=1}^2 Q_C  \, . 
\end{eqnarray}

\smallskip
\noindent
For $E_7 \ra su(6) \times u(1) \times su(2)$, with completely analogous
notation, we find  
\begin{equation}
(b_1,b_2, b_3, b_4, b_5) = ( a_7,  a_3, a_4, a_5,  a_6) ,  \quad c = a_1 \,,
\end{equation}
\begin{eqnarray}
 Q^* &=& 3a_1 +6a_2 +8a_3 +6a_4 +4a_5 +2a_6 +4a_7 \nonumber\\
    &=& -2\sum_{i=1}^6 Q_A^i -6Q_B +3\sum_{i=1}^2 Q_C\, . 
\end{eqnarray}

\smallskip
\noindent
Finally, for $E_8 \ra su(7) \times u(1) \times su(2)$ we have
\begin{equation}
(b_1,b_2, b_3, b_4, b_5, b_6) = ( a_8,  a_3,  a_4, a_5, a_6, a_7) , \quad c = a_1 \,,
\end{equation}
\begin{eqnarray}
Q^* &=& -14a_1 -28a_2 -40a_3 -32a_4 -24a_5 -16a_6 -8a_7 -
20a_8 \nonumber\\
    &=& 8\sum_{i=1}^6 Q_A^i +28Q_B -14\sum_{i=1}^2 Q_C\, . 
\end{eqnarray}

\end{document}